\documentclass[reprint,amsmath,amssymb,aps,superscriptaddress]{revtex4-2}

\usepackage{graphicx}
\usepackage{dcolumn}
\usepackage{amsmath}
\usepackage{bm}
\usepackage{mathrsfs}
\usepackage{txfonts}
\usepackage{epstopdf}
\usepackage{subfigure}
\usepackage{caption}
\usepackage{float}
\begin{document}
	
\title{Switching modulation of spin transport in ferromagnetic tetragonal silicene}
	
\author{Liehong Liao}
\affiliation{Department of Physics, Hangzhou Dianzi University, Hangzhou, Zhejiang 310018, China}

\author{Ying Ding}
\affiliation{Department of Physics, Hangzhou Dianzi University, Hangzhou, Zhejiang 310018, China}

\author{Fei Wan}
\affiliation{Department of Physics, Hangzhou Dianzi University, Hangzhou, Zhejiang 310018, China}

\author{Jiayan Zhang}
\affiliation{Department of Physics, Hangzhou Dianzi University, Hangzhou, Zhejiang 310018, China}

\author{Zhihui Chen}
\affiliation{Department of Physics, Hangzhou Dianzi University, Hangzhou, Zhejiang 310018, China}

\author{Xinyu Cheng}
\affiliation{Department of Physics, Hangzhou Dianzi University, Hangzhou, Zhejiang 310018, China}

\author{Ru Bai}
\email[]{bairu@hdu.edu.cn}
\affiliation{Center for Integrated Spintronic Devices (CISD), Hangzhou Dianzi University, Hangzhou, Zhejiang 310018, China}

\author{Gaofeng Xu}
\affiliation{Department of Physics, Hangzhou Dianzi University, Hangzhou, Zhejiang 310018, China}

\author{Yuan Li}
\email[]{liyuan@hdu.edu.cn}
\affiliation{Department of Physics, Hangzhou Dianzi University, Hangzhou, Zhejiang 310018, China}

\begin{abstract}
We study the band structure and transport properties of ferromagnetic tetragonal silicene nanoribbons by using the non-equilibrium Green's function method. The band structure and spin-dependent conductance are discussed under the combined effect of the external electric field, potential energy, exchange field and the spin-orbit coupling. One can easily realize a phase transition from a semimetallic to a semiconducting state by changing the transverse width of the nanoribbon. Separation of spin-dependent conductances arises from the effect of exchange field and the spin-orbit coupling, while zero-conductance behaviors exhibit spin-dependent band gaps induced by the electric field. We propose a device configuration of four-terminal tetragonal silicene nanoribbon with two central channels. It is found that spin current can be controlled by utilizing two switches. The switch with a high potential barrier can block electrons flowing from the central scattering region into other terminals. Interestingly, applying only one switch can realize spin-dependent zero conductance and large spin polarization. Two switches can provide multiple operations for controlling spin-dependent transport properties. The two-channel ferromagnetic tetragonal silicene nanoribbon can realize an effective separation of spin current, which may be a potential candidate for spintronic devices.
\end{abstract}
\date{\today}
\maketitle

\section{Introduction}
Spintronics involves the study of an effective manipulation of spin polarization in solid-state systems~\cite{Datta1990,Zutic2004,Hirohata2014}.
Two-dimensional hexagonal lattices with unique electronic and superior optoelectronic has triggered worldwide interest in the fields of condensed matter physics and device physics. The buckled two-dimensional crystal, such as silicene, germanene and stanene would enable new classes of nanoelectronic and spintronic devices~\cite{Molle2017}. Vertical stacking 2D materials can provide device applications in vertical transistors, infrared photodetectors, and spintronic transistors~\cite{Liang2019}. As a low-buckling hexagonal lattice composed of single-layer silicon atoms, silicene has many different properties compared with graphene. Due to $sp^3$ hybridization,  silicene is very unstable when exposed to air, so a substrate is needed for its epitaxial growth. In recent years, it has been synthesized on the metal surface, such as Ag (111) and Ir (111)~\cite{Ag111, Ir111}. Then silicene has aroused people great interest in theoretical and experimental researches~\cite{si1, si2, si3}. Silicene has a relatively large spin-orbit coupling (SOC), which can open a relatively large band gap near the Dirac point~\cite{gap}. Therefore, it has great potential applications in spintronics and valleytronics~\cite{Rycerz2007,Li2018}.

In the early days, it was considered that the hexagonal lattice is a necessary condition for the emergence of Dirac points.
Recently, the viewpoint that non-hexagonal honeycomb structures have stable Dirac cones has been proposed~\cite{4pssb2011,4PRL2012,4PRL,4Nano2014}. Researchers have many predictions of graphene, silicon carbide and silicene allotrope structures, and the tetragonal structure proves this viewpoint. Similar to tetragonal graphene~\cite{TG1994}, tetragonal silicene has strong stability near the Fermi level, and can maintain its structural stability at high temperature until $1000K$. Relevant studies have shown that this kind of silicene forms a buckling structure in space, which is a nodal line semimetal near the Fermi surface and has a linear dispersion relation. In addition, hydrogenation can induce a transition from the semimetal state to the semiconductor state for the tetragonal silicene. Some studies have shown that the band structure of tetragonal silicene can be modulated by adjusting the tight-binding parameters and on-site energies~\cite{MONDAL1}. Furthermore, it is proposed to modulate energy bands by utilizing the biaxial strain~\cite{Biaxial}. Some researchers have studied the optical and thermoelectric properties of tetragonal silicene by using the first-principle method~\cite{MONDAL2}, the results of which show highly anisotropy of its optical response. Hence, nanoribbons with different edge morphology can improve the optical and thermoelectric properties. What's more, tetragonal silicene was demonstrated to be a candidate for enhancing nonlinear optical and photocatalytic activities~\cite{optical2021}.

However, transport properties of tetragonal silicene have not been widely discussed. In this paper, the transport properties of tetragonal silicene heterojunction are calculated by using non-equilibrium Green's function method (NEGF). The spin orientation is separated by utilizing the combined effect of the electric field, exchange field and spin-orbit coupling, so as to achieve a perfect spin polarization. Our results show that the exchange field can break spin degeneracy and modulate energy bands in the presence of an electric field. Hence, the simultaneous action of the electric field, exchange field and spin-orbit coupling can help to achieve a large spin polarization. Furthermore, we design a four-terminal nanodevice with two-channel configurations. By applying opposite amplitudes of electric fields and exchange fields, spin-polarized currents can flow out of the two channels. Combining with the action of switches, we can easily obtain a desired spin-polarized current at the middle terminal.

In the study of the quantum transport, Dirac theory is a very effective method. It can help us to get analytical results and provide basic physical meaning of some problems. And for complex geometric structures, using a more general tight-binding model can more easily obtain complete band information. In addition, by utilizing NEGF method, the effects of lead and other interactions can be systematically incorporated into the tight-binding model~\cite{Wimmer}. It can deal with a wide range of conductors composed of scattering regions and external leads under the bias voltage. Therefore, compared with the effective Dirac Hamiltonian, the tight-binding NEGF technique has a wider range of applications. It is of great significance to develop and demonstrate its application in the tetragonal silicene systems.

The paper is organized as follows: In Sec.~\ref{sec:Model}, we give the theoretical calculating method of a tetragonal silicene nanoribbon under the influence of the spin-orbit coupling, the electric field and exchange field applied on the central scattering region. In Sec.~\ref{sec:Results}, we analyze the dispersion relations and employ NEGF to study the spin-dependent conductance. The effects of the external electric field, on-site energy, exchange field and the spin-orbit coupling on the spin-polarized current are analyzed and discussed. In addition, motivated by the perfect spin polarization, we design a nanodevice to obtain spin-polarized currents. Finally, a summary is given in Sec.~\ref{sec:Conclusions}.
\section{THEORETICAL MODEL}\label{sec:Model}
\begin{figure}[t]
	\centering
	\includegraphics[scale=0.35,trim=105 100 0 20,clip]{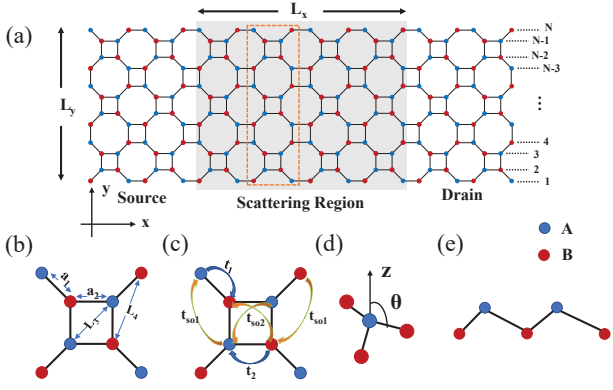}%
	
	\caption{(a) The schematic of tetragonal silicene heterojunction. (b) The lattice structure of the unit cell of tetragonal silicene nanoribbon. (c) The nearest-neighbor hopping terms $t_1$, $t_2$, and the next-nearest-neighbor hopping terms $t_{so1}$, $t_{so2}$ [SOC]. (d) z-axis is perpendicular to the plane, and $\theta$ is defined as the angle between the Si-Si bond and z-axis. (e) Side view of the low-buckled tetragonal silicene structure.}
	\label{fig:lead}
\end{figure}
We consider the dynamics of electrons hopping in the tetragonal lattice governed by the nearest-neighbor tight-binding Hamiltonian~\cite{gap,MONDAL1,SOC1}
\begin{eqnarray}\label{eq:parameter}
\begin{aligned}
	H=\sum_{i\alpha}V_ic^\dag_{i\alpha} c_{i\alpha}+\sum_{\langle i,j\rangle,\alpha}(t_1c^\dag_{i\alpha}c_{j\alpha}+t_2c^\dag_{i\alpha}c_{j\alpha})
\\+\sum_{i\alpha}\mu_ia_zE_zc^\dag_{i\alpha} c_{i\alpha}+\sum_{i\alpha}\ Ms_zc^\dag_{i\alpha} c_{i\alpha}.
\end{aligned}
\end{eqnarray}

The first term represents the on-site energy adjusted by top or bottom gates, $c_{i,\alpha}~(c^\dag_{i,\alpha})$ represents the spin $\alpha$ electron annihilation (creation) operator on the site $i$. The second term is the nearest-neighbor hopping between lattice sites, the summation $\sum_{\langle i,j\rangle} $ runs over the nearest-neighbor sites, $t_1$ and $t_2$ are hopping energies with different bond length $a_1$ and $a_2$, respectively, which are shown in Fig.~\ref{fig:lead}(b)-(c). The third term describes the contribution of the sublattice potential caused by the vertical electric field, $a_z=0.49\AA$ being the distance of the two sublattice planes. Where $\mu_i=\pm1$ for $A(B)$ site, and z axis is shown in Fig.~\ref{fig:lead}(d). For convenience, we use $E_{\xi}=a_z E_z$ to denote the amplitude of the sublattice potential. The fourth term describes the contribution of the ferromagnetic exchange field, with $M$ being the amplitude of the exchange field and $s_z=\pm 1$. The exchange field may be achieved via magnetic proximity coupling to a ferromagnet such as depositing Fe atoms to the tetragonal silicene surface or depositing tetragonal silicene to a ferromagnetic insulating substrate~\cite{Yang2013,Wei2016,An2012,Missault2015}.
What's more, the side view of the low-buckled tetragonal silicene structure is plotted in Fig.~\ref{fig:lead}(e). Adopting the optimized parameters obtained from DFT method~\cite{MONDAL1}, we set the bond lengths $a_1 = 2.252\AA$ and $a_2 = 2.304\AA$. The relaxed equilibrium values for the hopping parameters are $t_1 = 0.85 eV$ and $t_2 = 0.90 eV$. Note that we first ignore the spin-orbit couplings, namely the next-nearest-neighboring hopping terms.

We devise a nanoribbon of tetragonal silicene with an armchair-edge configuration (ATSNR). As shown in Fig.~\ref{fig:lead}(a), it consists of the central scattering region, left (source) and right (drain) leads. Where the orange dashed rectangle indicates the smallest repeating unit along the transport direction. $N$ denotes the site number along y direction, $L_x$ and $L_y$ represent the length and width of the scattering region. It is expected that the transport properties of the system can be regulated by changing the spin-orbit coupling and the external conditions of the central scattering region.
\begin{figure}[b]
	\centering
	\includegraphics[scale=0.28,trim=0 0 0 0,clip]{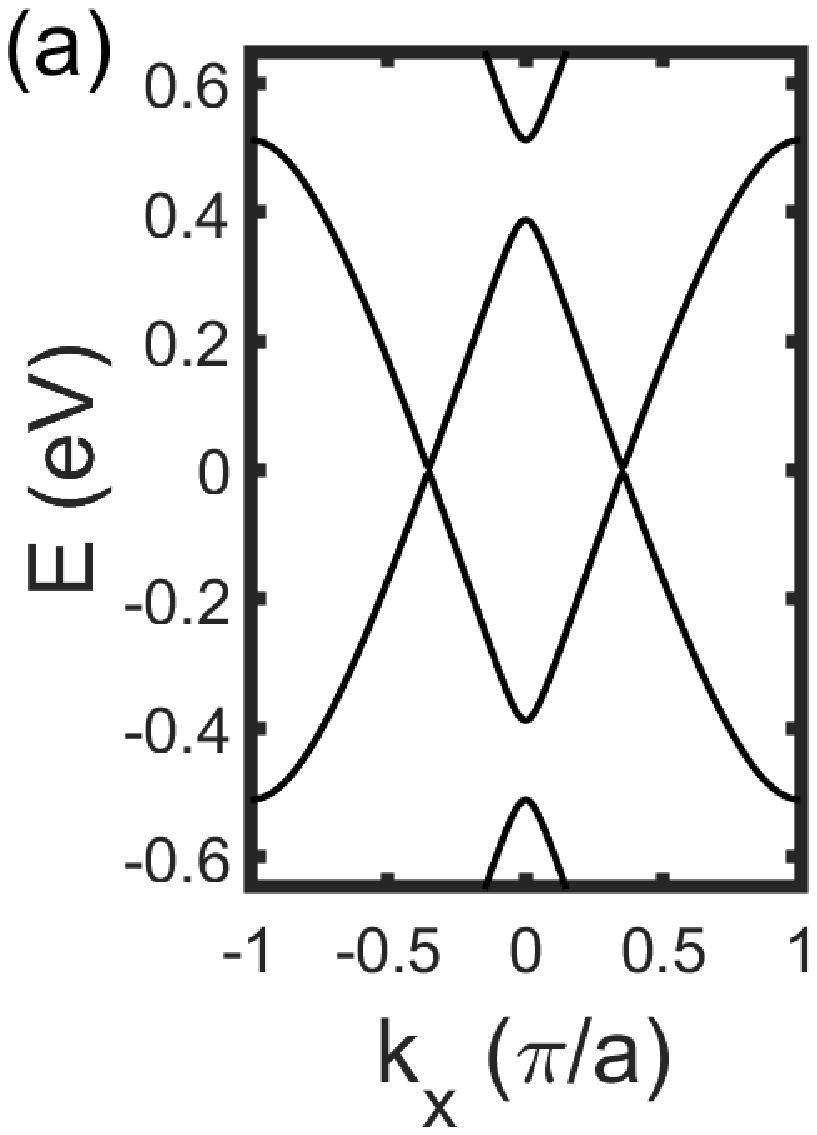}%
	\includegraphics[scale=0.28,trim=0 0 0 0,clip]{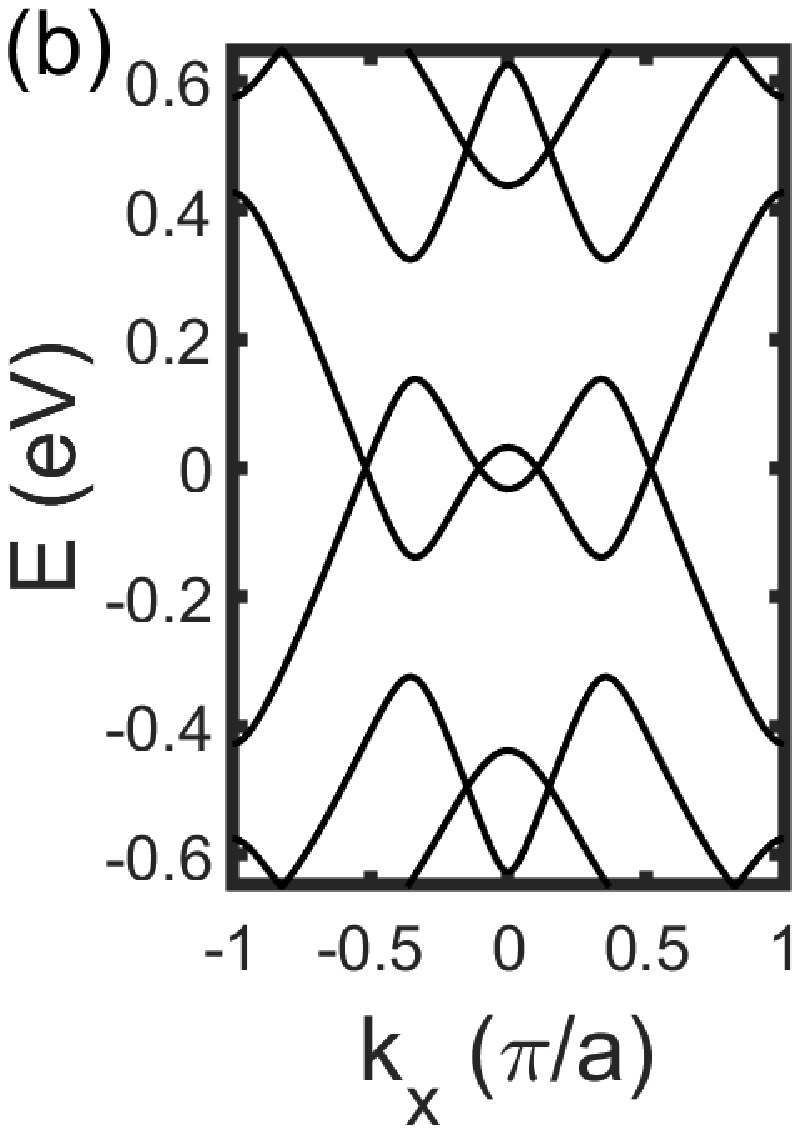}%
    \includegraphics[scale=0.28,trim=0 0 0 0,clip]{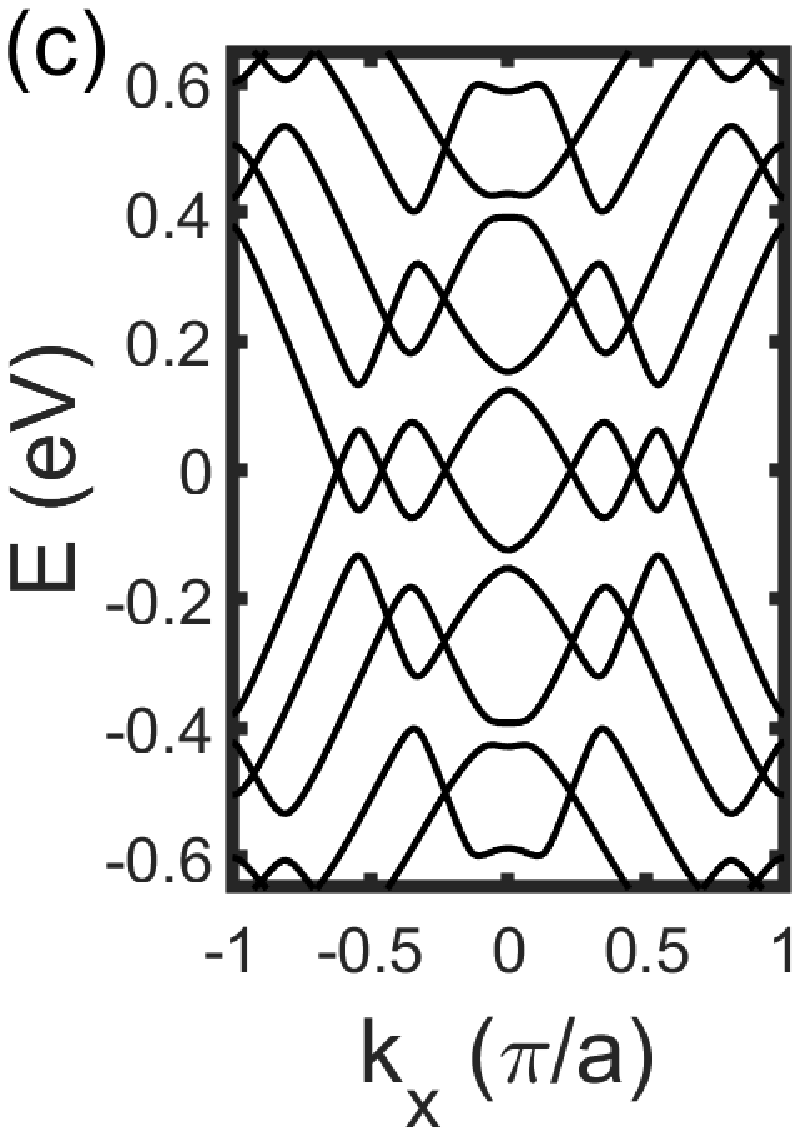}%

	\includegraphics[scale=0.28,trim=0 0 0 0,clip]{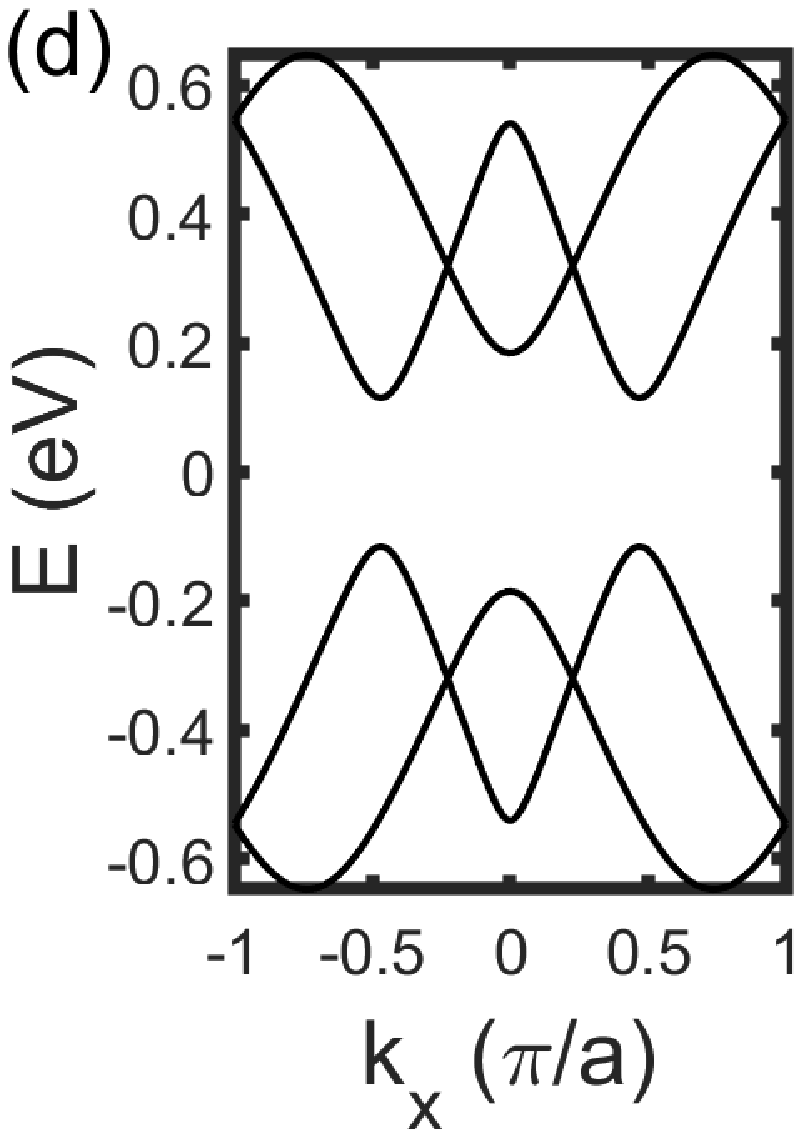}%
    \includegraphics[scale=0.28,trim=0 0 0 0,clip]{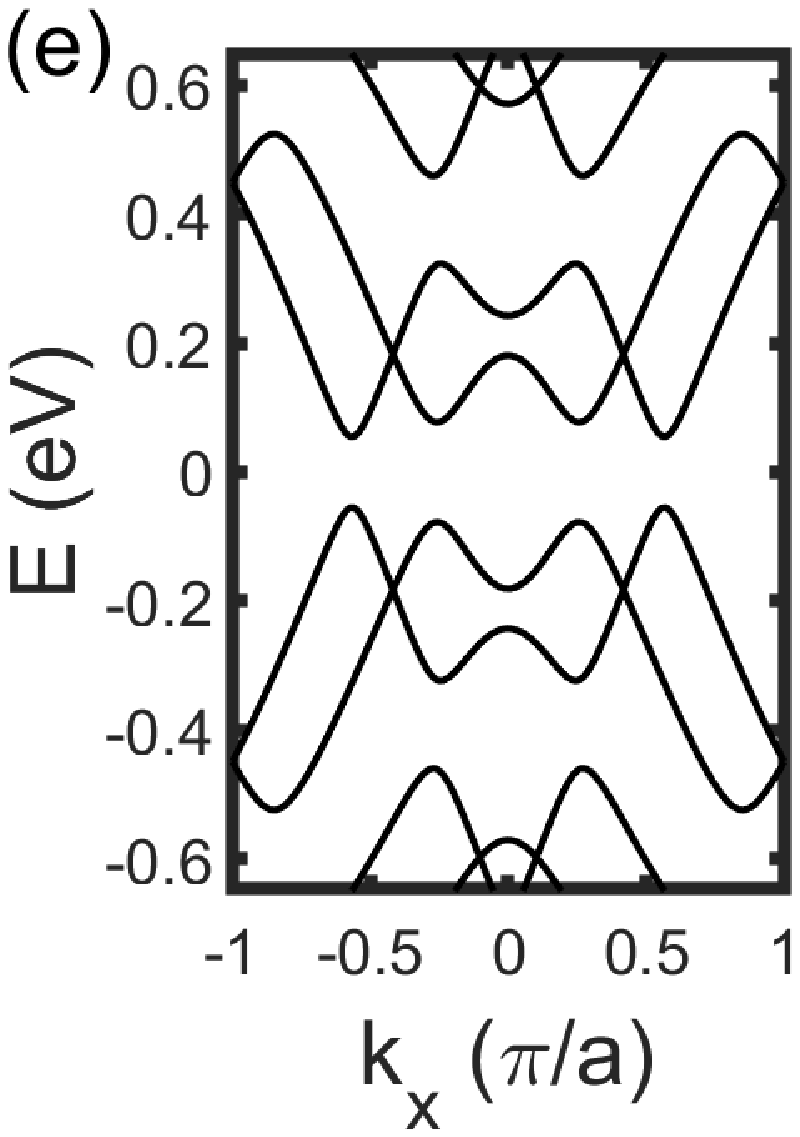}%
    \includegraphics[scale=0.28,trim=0 0 0 0,clip]{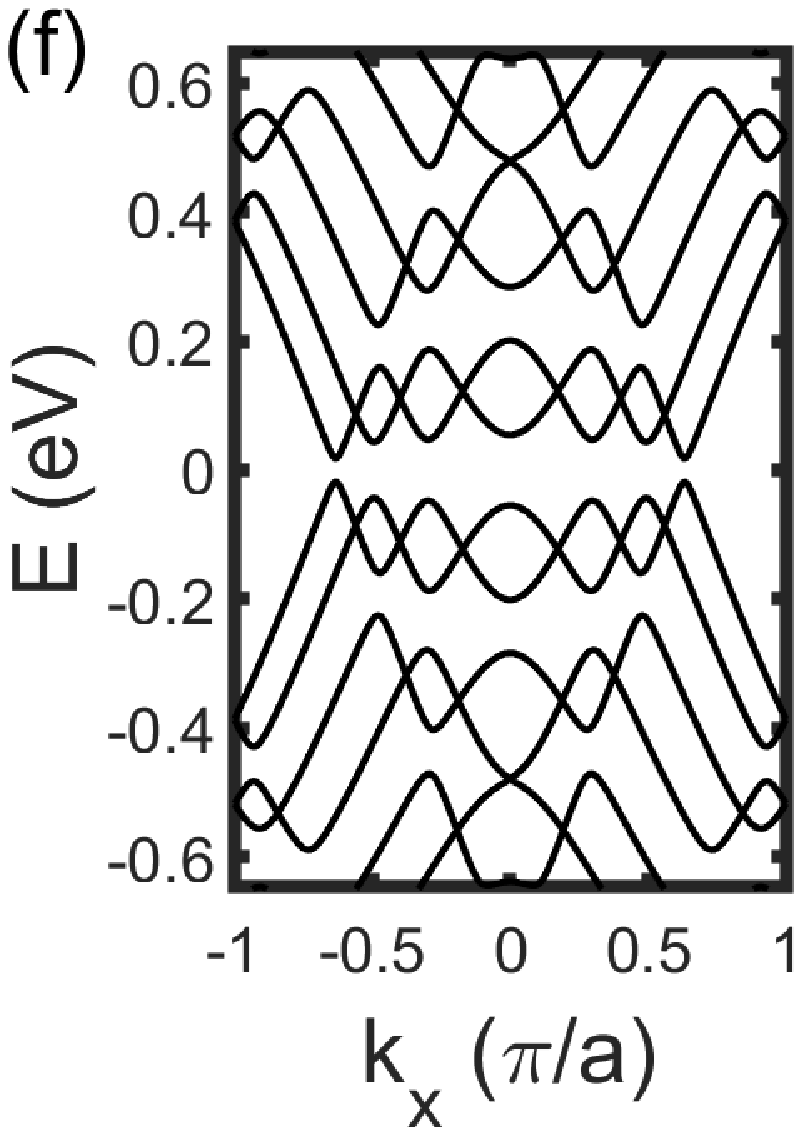}%
	\caption{Dispersion relations plotted as a function of the wave vector $k_x$ for different values of $N_y$ with (a) $N_y=4$, (b) $N_y=8$, (c) $N_y=16$, (d) $N_y=6$, (e)$N_y=10$, (f) $N_y=18$.}
	\label{fig:band0}
\end{figure}

One can calculate the conductance of the tetragonal silicene nanoribbon in terms of Landauer-B\"{u}ttiker formula~\cite{G1985,Nikolic,Wan2021},
 \begin{equation}\label{eq:Cond}
	G=G_0\mathrm{T}=\frac{e^2}{h}\left(
\begin{array}{ccc}
T^{\uparrow\uparrow} & T^{\uparrow\downarrow} \\
T^{\downarrow\uparrow} & T^{\downarrow\downarrow}
 \end{array}
 \right),
\end{equation}
here $G_0=e^2/h$, $T^{\alpha \beta}$ refers to the transmission probability to detect an electron with spin $\alpha$ in the drain lead arising from an injected electron with spin $\beta$ in the source lead. The transmission coefficient can be calculated as $\mathrm{T^{\alpha \beta}}=\mathrm{Tr}[\Gamma^\alpha_L G^r \Gamma_R^{\beta} G^a]$, where $\Gamma^\alpha_{L/R}=i(\Sigma_{L/R}^{r,\alpha}-\Sigma_{L/R}^{a,\alpha})$ are line-width functions with the retarded/advanced self-energy $\Sigma_{L/R}^{r/a,\alpha}$ for spin channel $\alpha$.
After numerically calculating and solving the surface Green's function, one can easily obtain the self energy giving the coupling effect of the two leads. Then we can calculate the Green's function of the system
 \begin{eqnarray}
	G^r={[G^a]}^\dag={[E\mathrm{I}+i\eta-H_c-{\Sigma}_L^r-{\Sigma}_R^r]}^{-1},
\end{eqnarray}
where $H_c$ represents the Hamiltonian of the central scattering region, $E$ is Feimi energy and $\mathrm{I}$ is identity matrix,
the self-energy matrix can be written as
 \begin{eqnarray}
	\Sigma_{L,R}^r=\left(
\begin{array}{ccc}
\Sigma_{L,R}^{r,\uparrow} & 0 \\
0 & \Sigma_{L,R}^{r,\downarrow}
 \end{array}
 \right).
\end{eqnarray}

\section{NUMERICAL RESULTS AND DISCUSSION}\label{sec:Results}
\begin{figure}[t]
	\centering
	\includegraphics[scale=0.38,trim=0 0 0 0,clip]{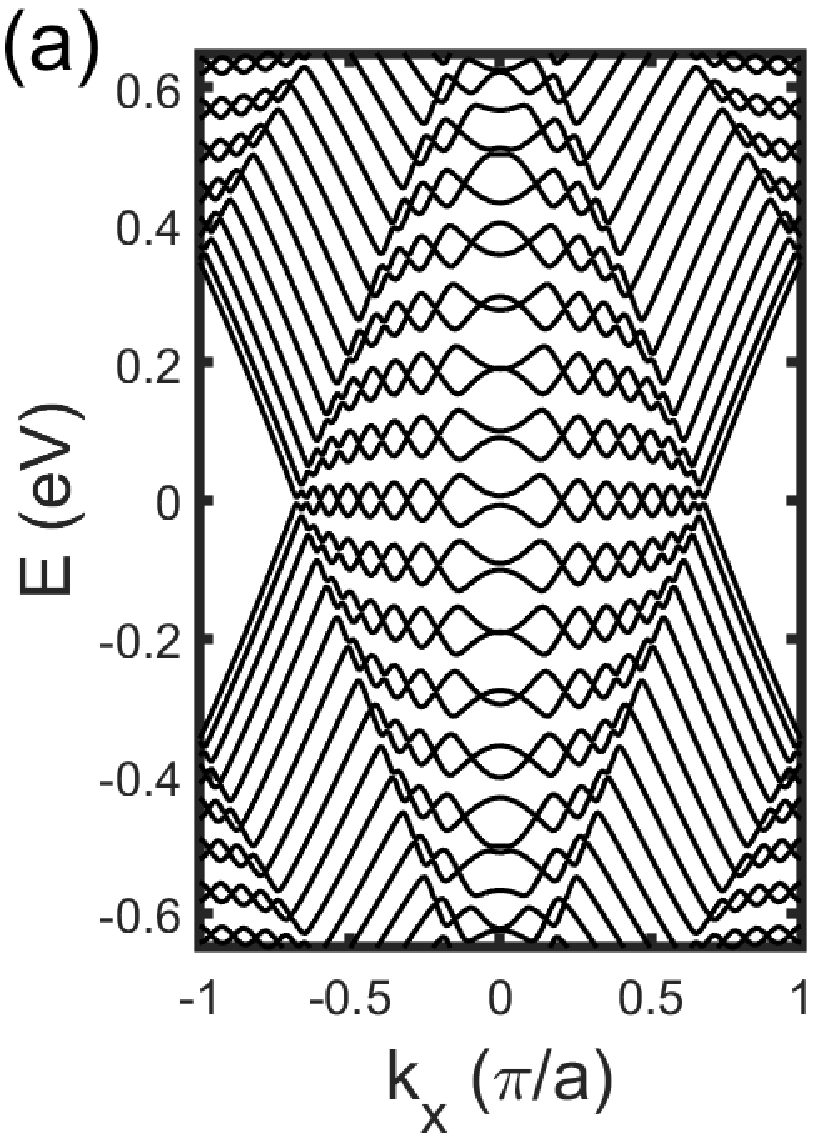}%
	\includegraphics[scale=0.38,trim=0 0 0 0,clip]{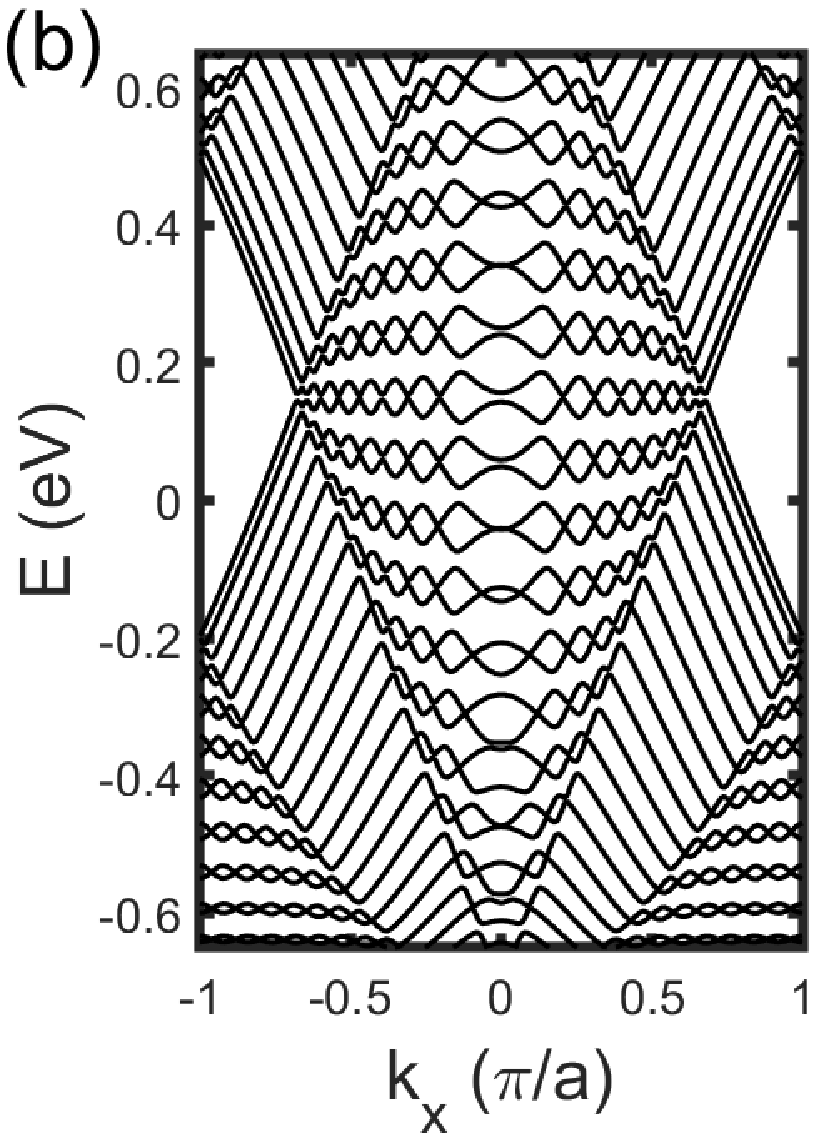}%

    \includegraphics[scale=0.38,trim=0 0 0 0,clip]{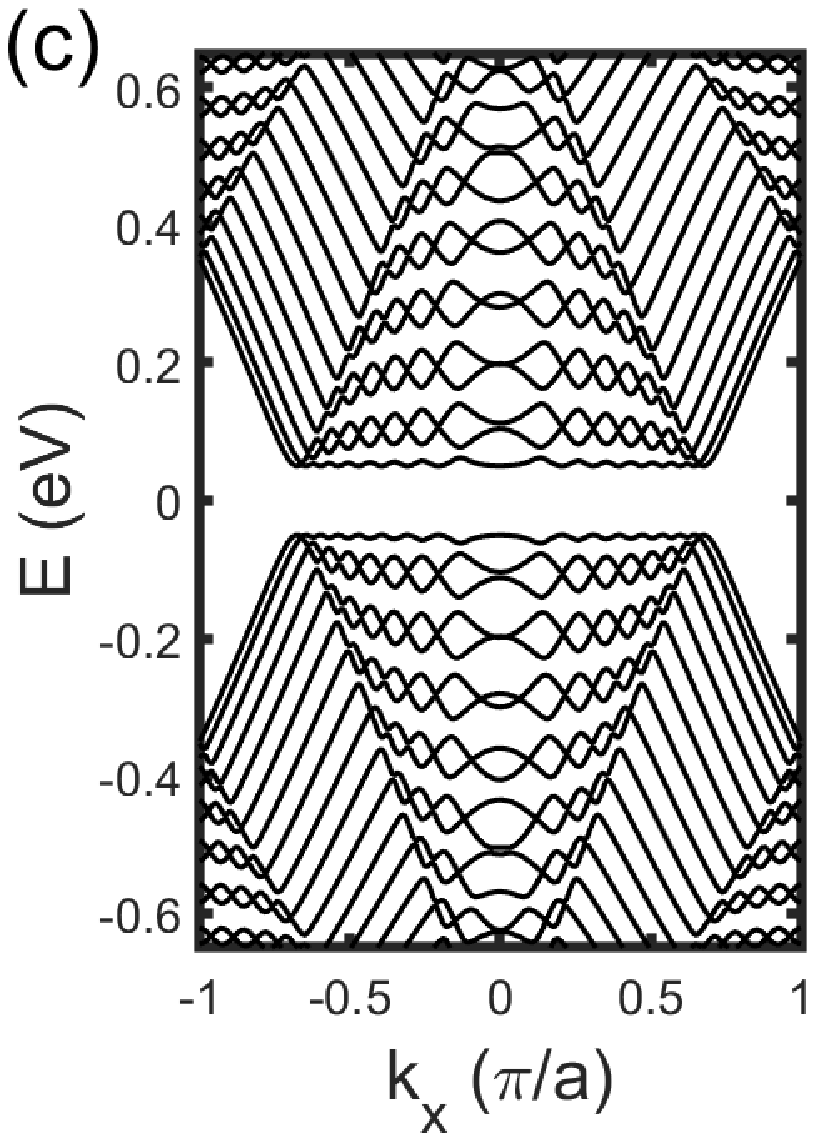}%
	\includegraphics[scale=0.38,trim=0 0 0 0,clip]{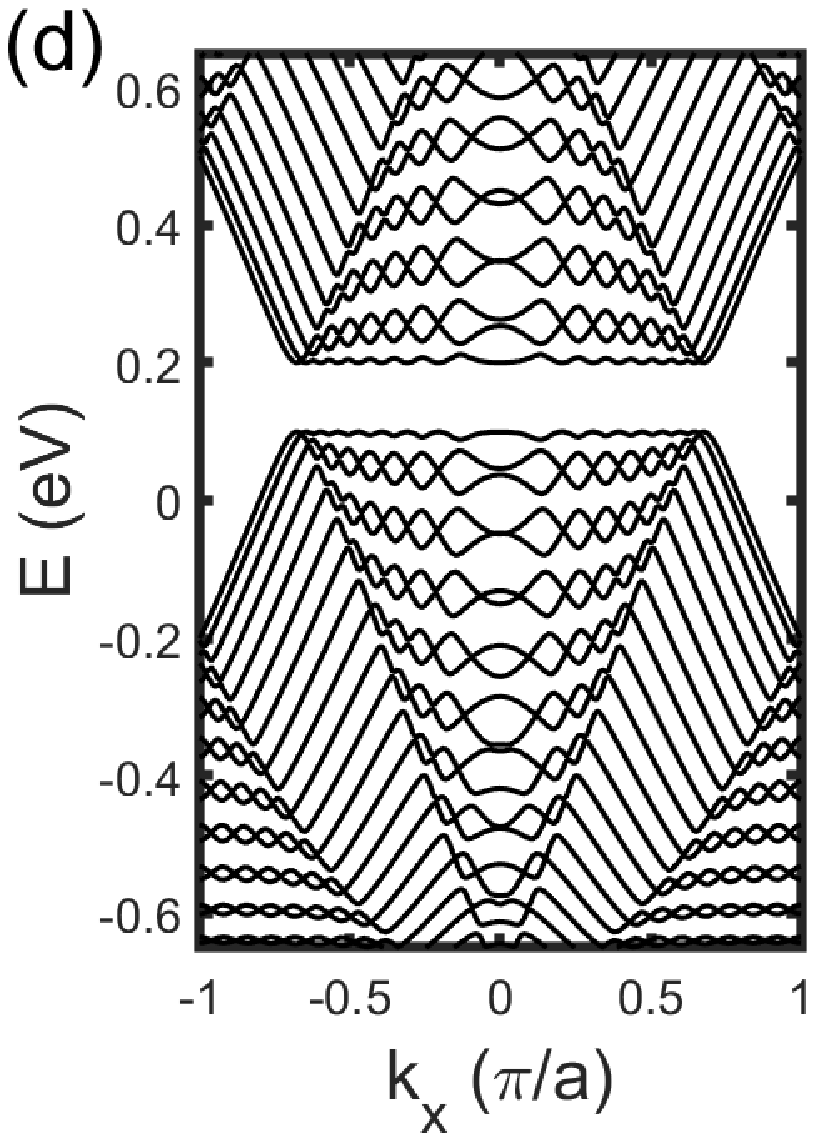}%
\caption{Band structures of ATSNR plotted as a function of the wave vector $k_x$ for different on-site energies and sublattice potentials (a) $V_0=0$ and $E_\xi=0$; (b) $V_0=0.15$ eV and $E_\xi=0$; (c) $V_0=0$ and $E_\xi=0.05$ eV; (d) $V_0=0.15$ eV and $E_\xi=0.05$ eV. Other parameters are $N_x=280$ and $N_y=48$.}
	\label{fig:band}
\end{figure}

In Fig.~\ref{fig:lead}(a) for ATSNR, each minimum repeating unit describing by the orange dash rectangle contains two silicon chains.
The length and width of the central region are $L_x=(\sqrt{2}a_1+a_2)\times N_x/2+a_2\times(N_x/2-1)$ and $L_y=(\sqrt{2}a_1+a_2)\times (N_y/4)+a_2\times (N_y/4-1)$, respectively. Furthermore, the length and width of the nanoribbon can be represented by the site numbers of $N_x$ and $N_y$ along x and y direction. The size of the central region is chosen to be $L_x=127$ nm and $L_y=10.6$ nm, corresponding to $N_x = 280$ and $N_y = 48$, respectively.

As shown in Fig.~\ref{fig:band0}, the dispersion relation of ATSNR depends on the transverse width $N_y$ of the nanoribbon. Interestingly, there exist repeated phase transitions between semimetal states and semiconductor states when the site number changes from $N_y=4$ to $18$. In Figs.~\ref{fig:band0}(a)-\ref{fig:band0}(c),
we can see that the conduction bands near zero energy overlap with the valence bands and form a semimetal state when $N_y$ is a multiple of four, namely $N_y=4 n$ with the integer $n=1,2,3,\cdots$. The linear dispersion relation near zero energy in Fig.~\ref{fig:band0}(a) is consistent with that reported in Ref.~\cite{MONDAL1}.

Seen from Fig.~\ref{fig:lead}(a), when $N_y=4n$, there exist complete silicon rings, or a symmetrical axis along $x$ direction, thus the silicon atoms are symmetrical about the transverse symmetrical axis. It means that the nanoribbon with a transverse symmetry becomes a semimetalic system.
In Fig.~\ref{fig:band0}(d)-\ref{fig:band0}(f), $N_y=6,10,18$ which is not the multiple of four, there exists no transverse symmetrical axis. When $N_y=6$, a band gap about $240$ meV is opened and the system demonstrates a semiconducting state. With $N_y$ increasing from 6 to 18, the system still keeps a semiconducting state, but the band gap significantly decreases to about $20$ meV and more subbands occur. Obviously, one can easily realize a phase transition from a semimetalic state to a semiconducting state by changing the transverse width of the nanoribbon.

\subsection{Combined effect of the on-site energy and the electric field}
Next, we study the combined effect of the on-site energy and the electric field on energy bands and the system's conductance. In Fig.~\ref{fig:band}(a), there have multiple Dirac cones near zero energy and overlapped subbands, which show the semimetallicity when $V_0=0$ and $E_\xi=0$.
When the on-site energy $V_0=0.15$ eV is applied on the central region, the whole band structure moves up 0.15 eV.
The applied electric field can open a band gap about 0.1 eV, as shown in Fig.~\ref{fig:band}(c) and Fig.~\ref{fig:band}(d). We can find that the band gap and the positions of subbands can be modulated by the electric field and the on-site energy, while the overlapped structures of subbands keep unchanged.
\begin{figure}[t]
	\centering
	\includegraphics[scale=0.35,trim=0 0 0 0,clip]{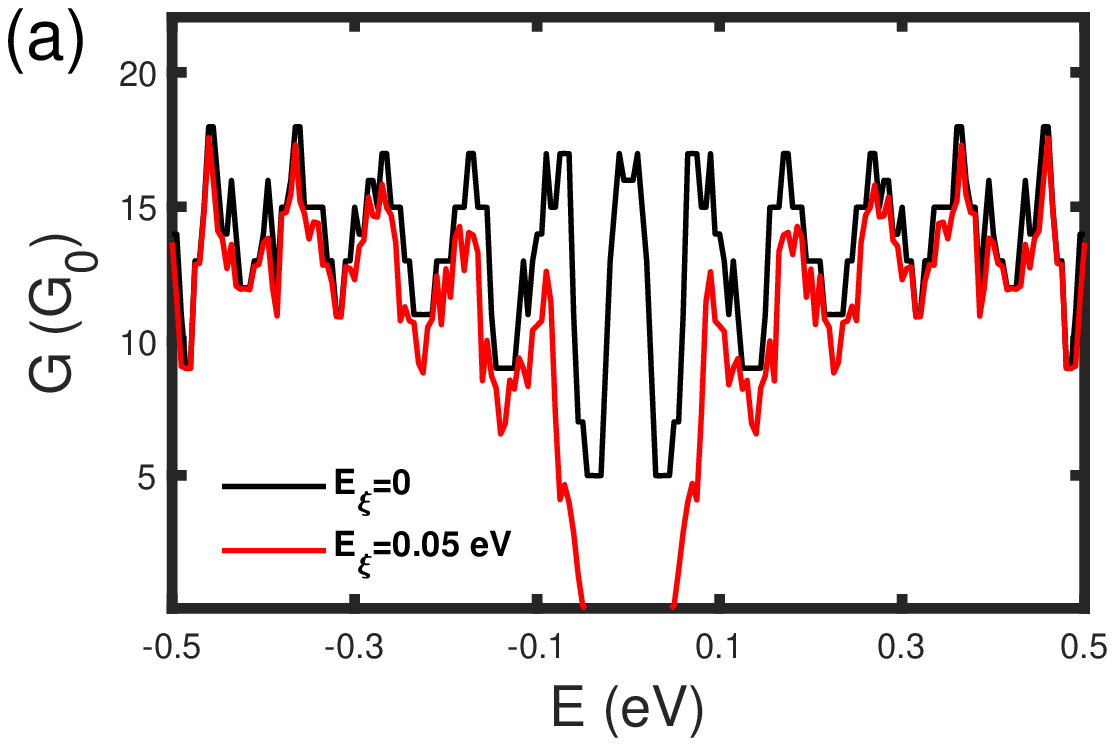}%
	\includegraphics[scale=0.35,trim=0 0 0 0,clip]{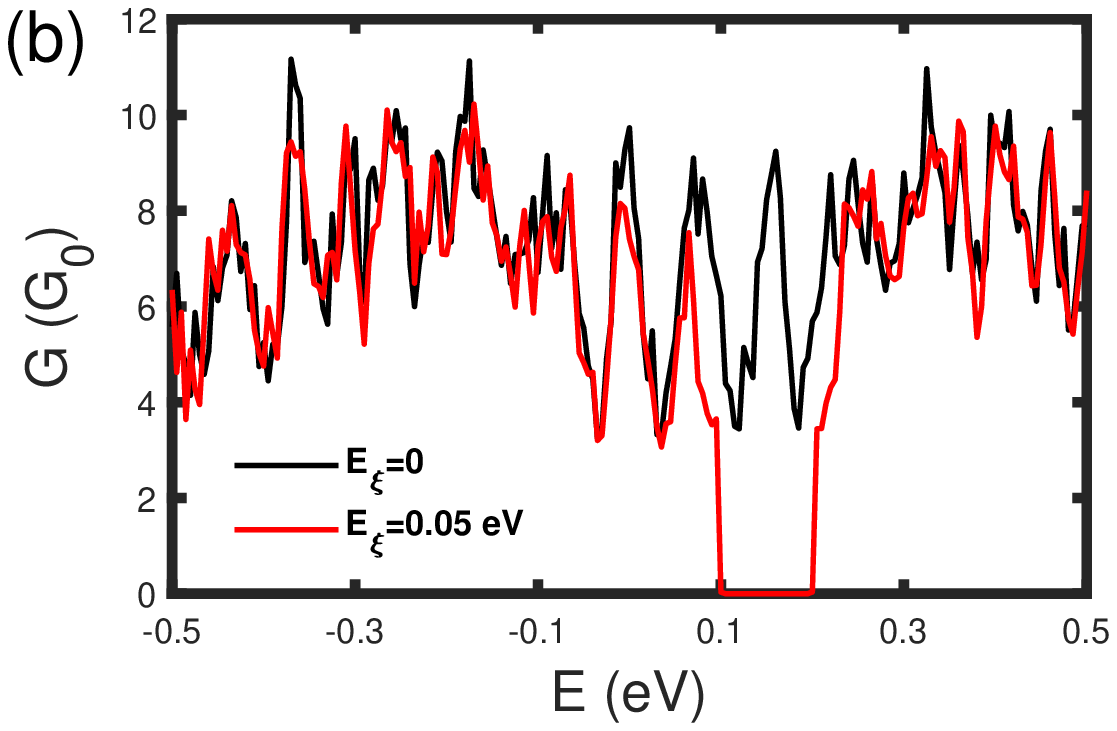}%
	\caption{The conductance $G$ plotted as a function of Fermi energy $E$ for on-site energies (a) $V_0=0$ and (b) $V=0.15$ eV. Other parameters are the same as those in Fig.~\ref{fig:band}.}
	\label{fig:G0}
\end{figure}

Therefore, one can open a band gap of the tetragonal silicene with the help of an external electric field to further inhibit the transmission of electrons and act as a nanoswitch. This idea is proved by our calculating results of transport properties. As shown in Fig. \ref{fig:G0}, the conductance of the tetragonal silicene was plotted as a function of Fermi energy for different parameters of external fields. In Fig. \ref{fig:G0}(a), the maximum value of the conductance is about $18G_0$, while the minimum value of the conductance is about $5 G_0$ when $E_\xi=0$. The conductance shows oscillating behaviors which reflecting the repeated and overlapped subbands in the wave vector regime $-0.65\pi/a<k_x<0.65 \pi/a$. Especially, there exists a conductance gap in the energy regime $-0.05 \mathrm{eV}<E<0.05\mathrm{eV}$ due to the effect of the electric field. When the on-site energy is $V_0=0.15$eV, the conductance gap moves right 0.15eV and also has oscillating behaviors [see Fig. \ref{fig:G0}(b)]. Correspondingly, the maximum value of the conductance decreases to about $11 G_0$.

\subsection{Combined effect of the exchange field and the electric field}
\begin{figure}[t]
	\centering
	\includegraphics[scale=0.38,trim=0 0 0 0,clip]{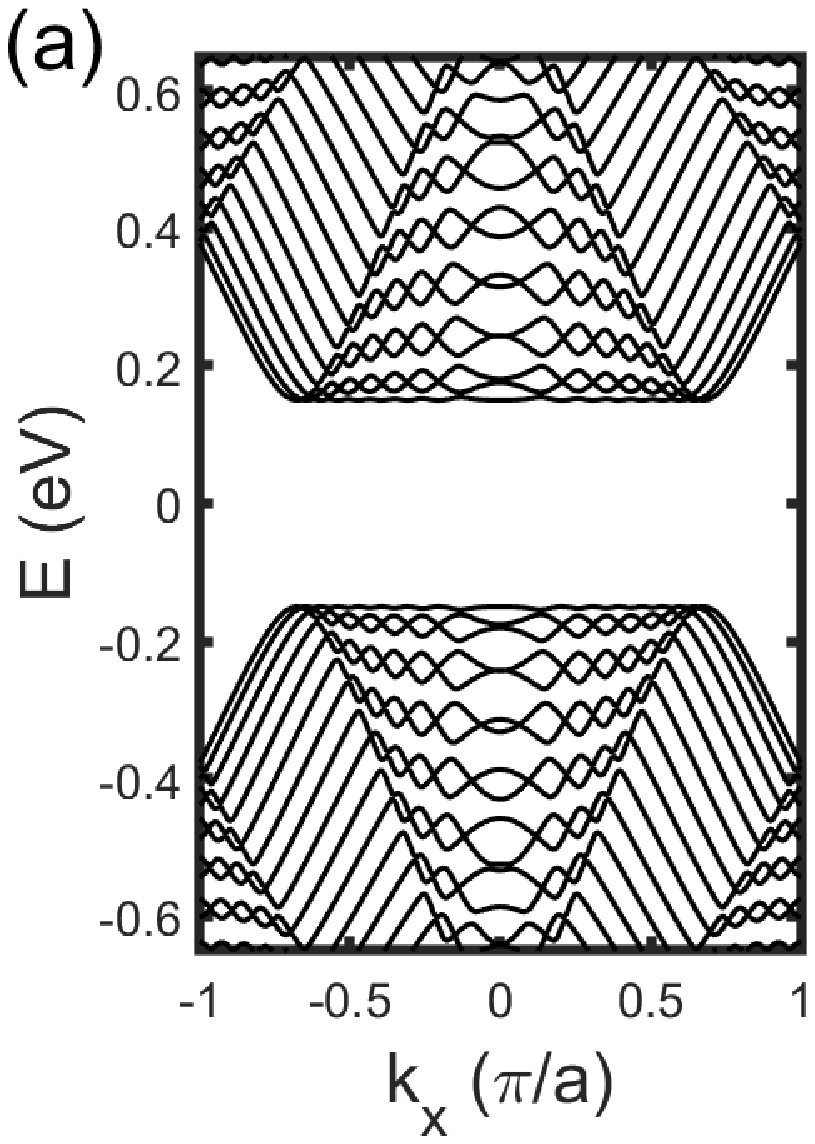}%
    \includegraphics[scale=0.38,trim=0 0 0 0,clip]{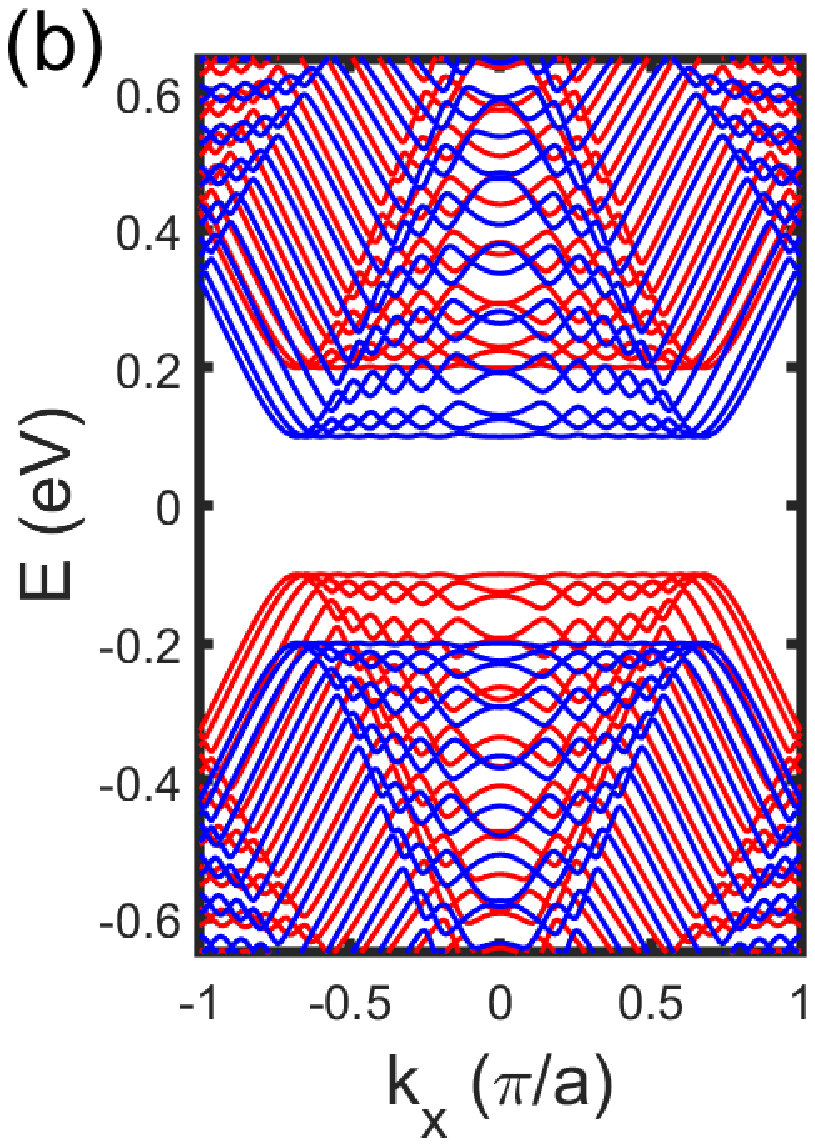}%

    \includegraphics[scale=0.38,trim=0 0 0 0,clip]{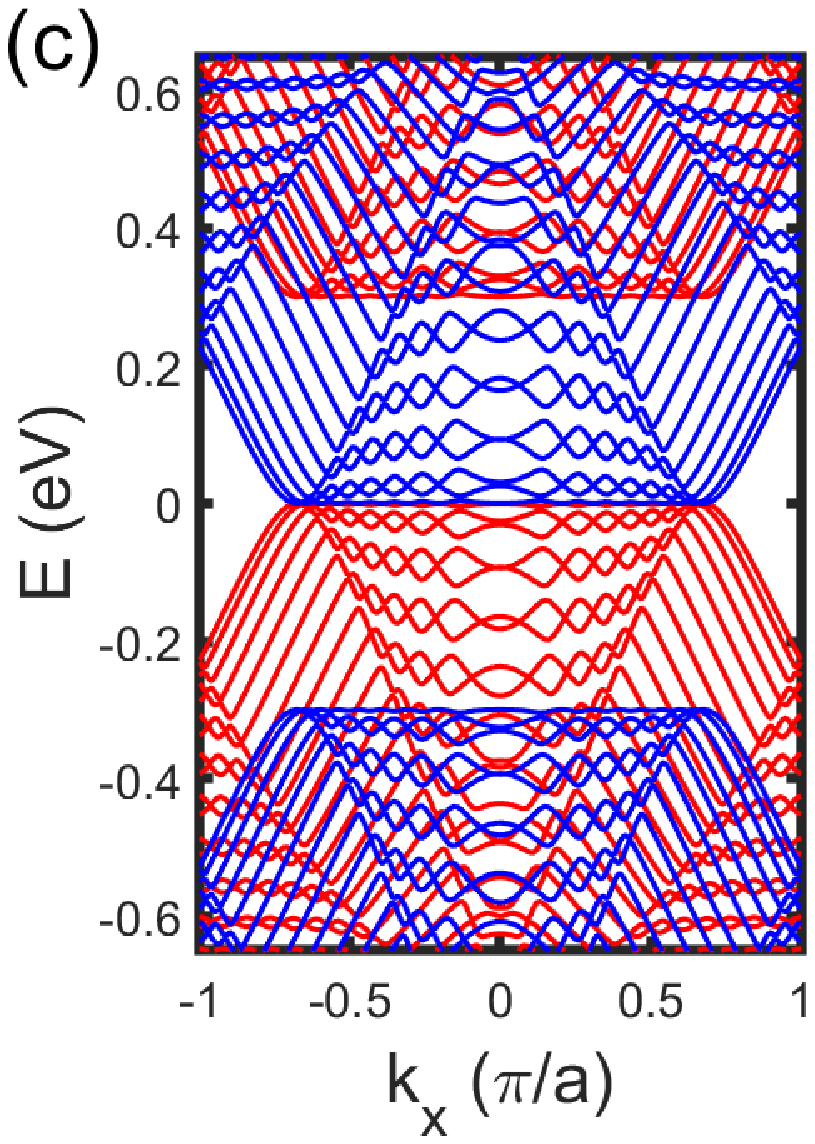}%
    \includegraphics[scale=0.38,trim=0 0 0 0,clip]{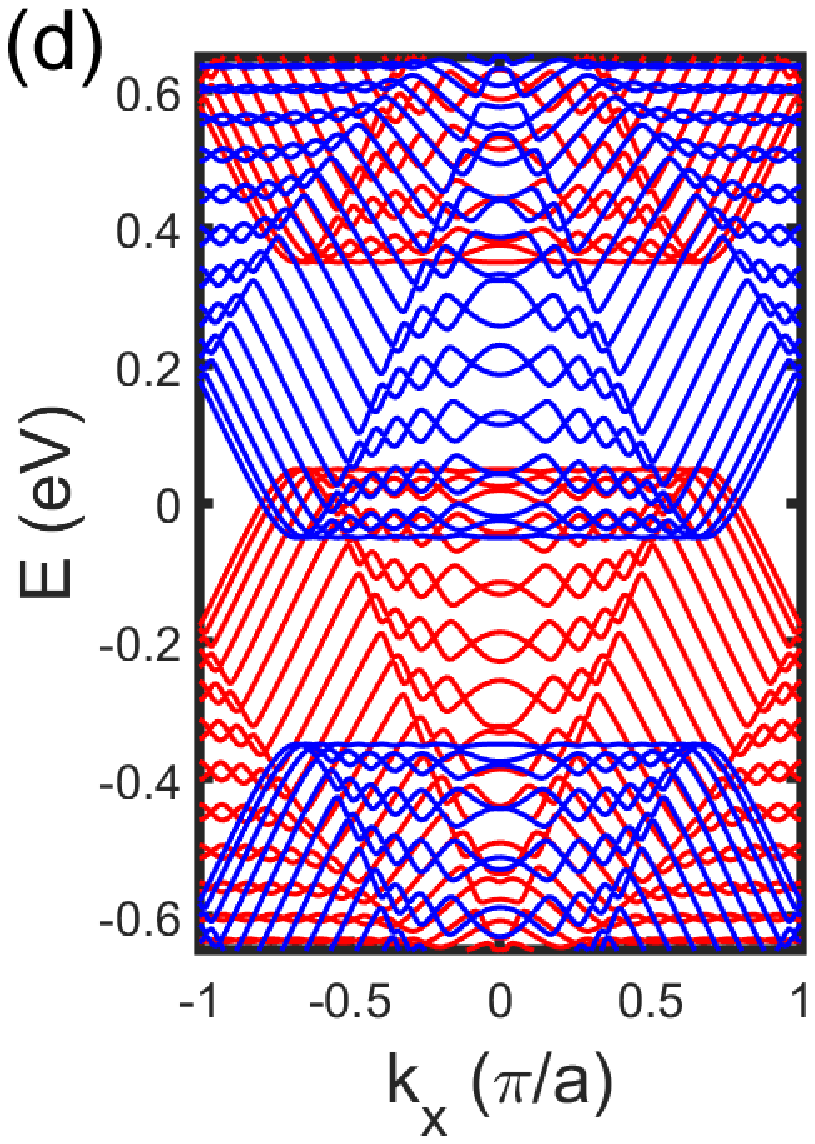}%
	\caption{Spin-dependent energy bands of ATSNR for different exchange fields (a) $M=0$, (b)$M=0.05$ eV, (c) $M=0.15$ eV, (d) $M=0.2$ eV. Red (blue) curves represent spin-up (spin-down) energy bands, and black curves represent total energy bands with spin degeneracy. The sublattice potential is $E_\xi=0.15$ eV, other parameters are the same as those in Fig.~\ref{fig:band}.}
	\label{fig:spinband1}
\end{figure}
In recent decades, the proposed concept of spintronics has brought new applications to the design of spin-based electronic devices~\cite{Zutic2004,spindevice}. It is crucial for spin-based devices to realize an effective modulation of spin polarization. The exchange field can be useful for improving the spin polarization of the system. Thus we further study the combined effect of the exchange field and the electric field on the energy bands in ATSNR.

First, we fix the sublattice potenial to $E_\xi=0.15$ eV. In Fig.~\ref{fig:spinband1} we plot the dispersion relation as a function of the wave vector $k_x$. When $M=0$, we can see that the energy bands are spin degeneracy, and there exists a band gap with the amplitude of $0.3 eV$. With the exchange field increasing to $M=0.05$ eV, the spin-up energy bands move up 0.05 eV, while spin-down energy bands move downwards. The band gap correspondingly decreases to about 0.2 eV due to the opposite shift of spin-dependent energy bands. When $M=0.15$ eV, the spin-down conduction band touches with the spin-up valence band, thus resulting in disappearance of the band gap. With the amplitude of the exchange field increasing to 0.2 eV, spin-dependent energy bands overlap with each other, and the maximum value of the spin-up valence band is larger than the minimum value of the spin-down conduction band [see Fig. \ref{fig:spinband1}(d)].

\begin{figure}[t]
	\centering
	\includegraphics[scale=0.38,trim=0 0 0 0,clip]{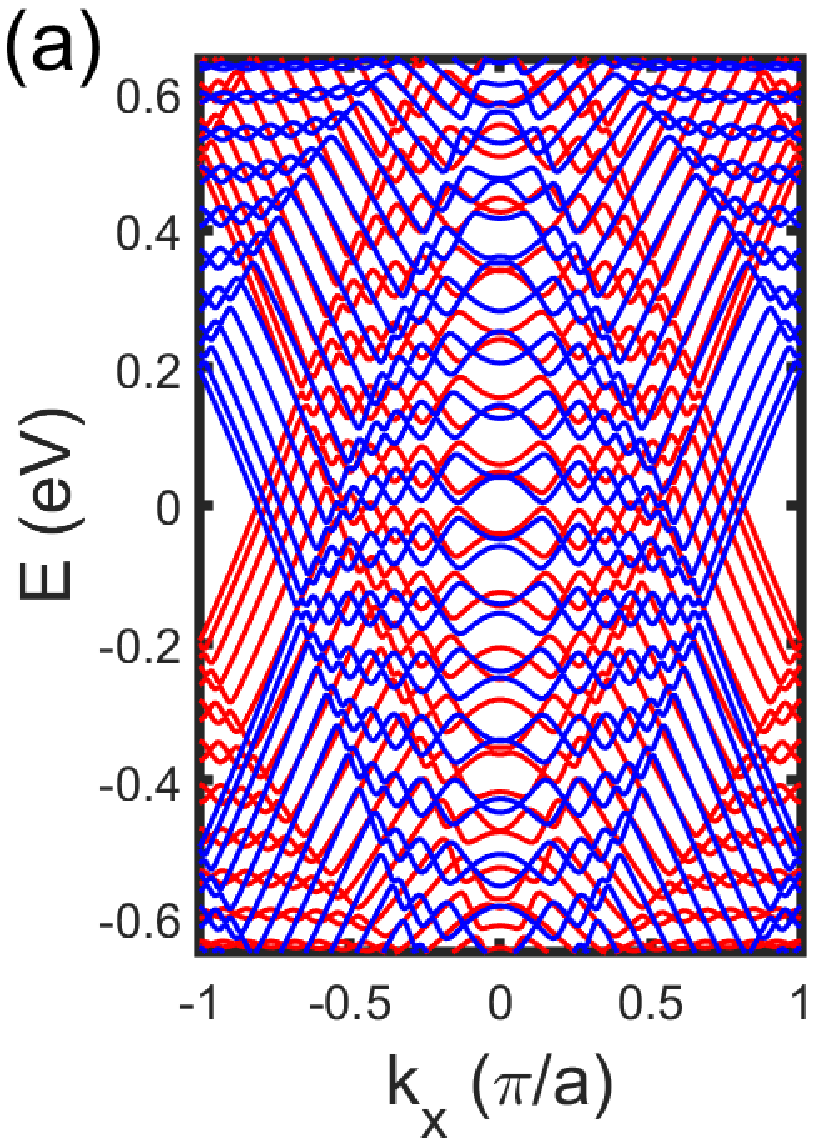}%
    \includegraphics[scale=0.38,trim=0 0 0 0,clip]{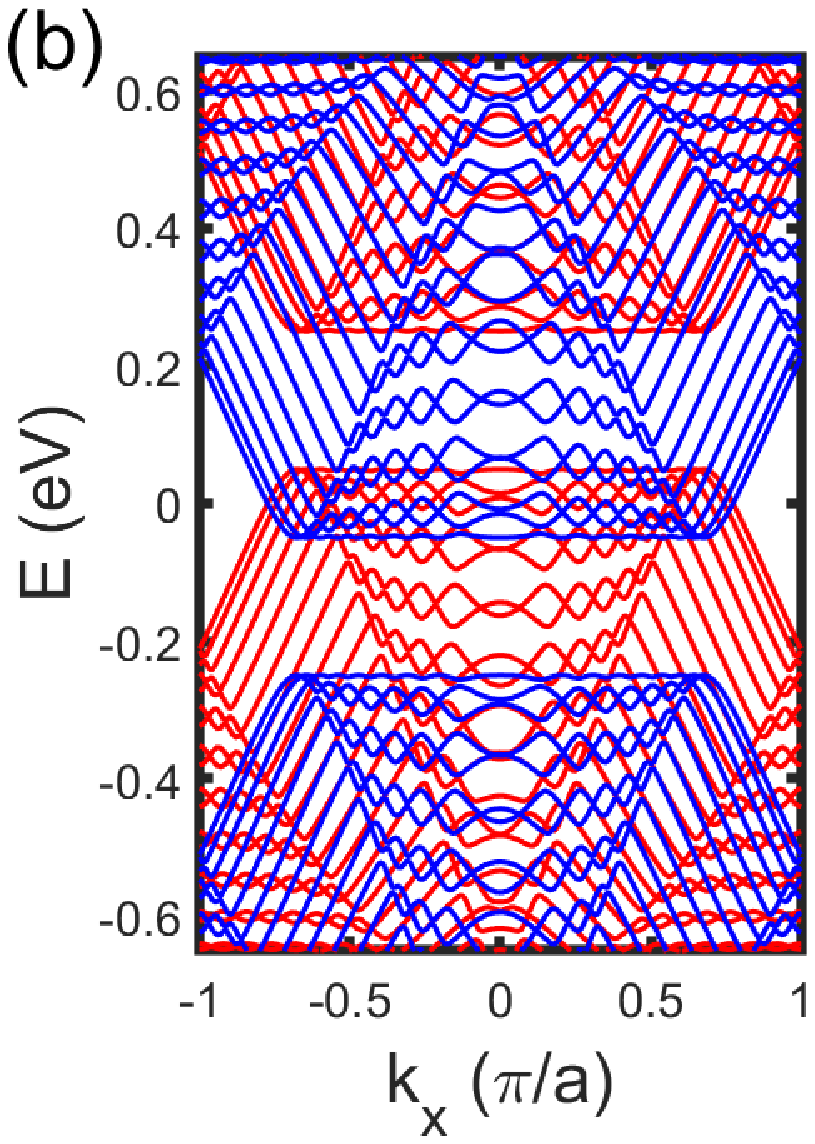}%

    \includegraphics[scale=0.38,trim=0 0 0 0,clip]{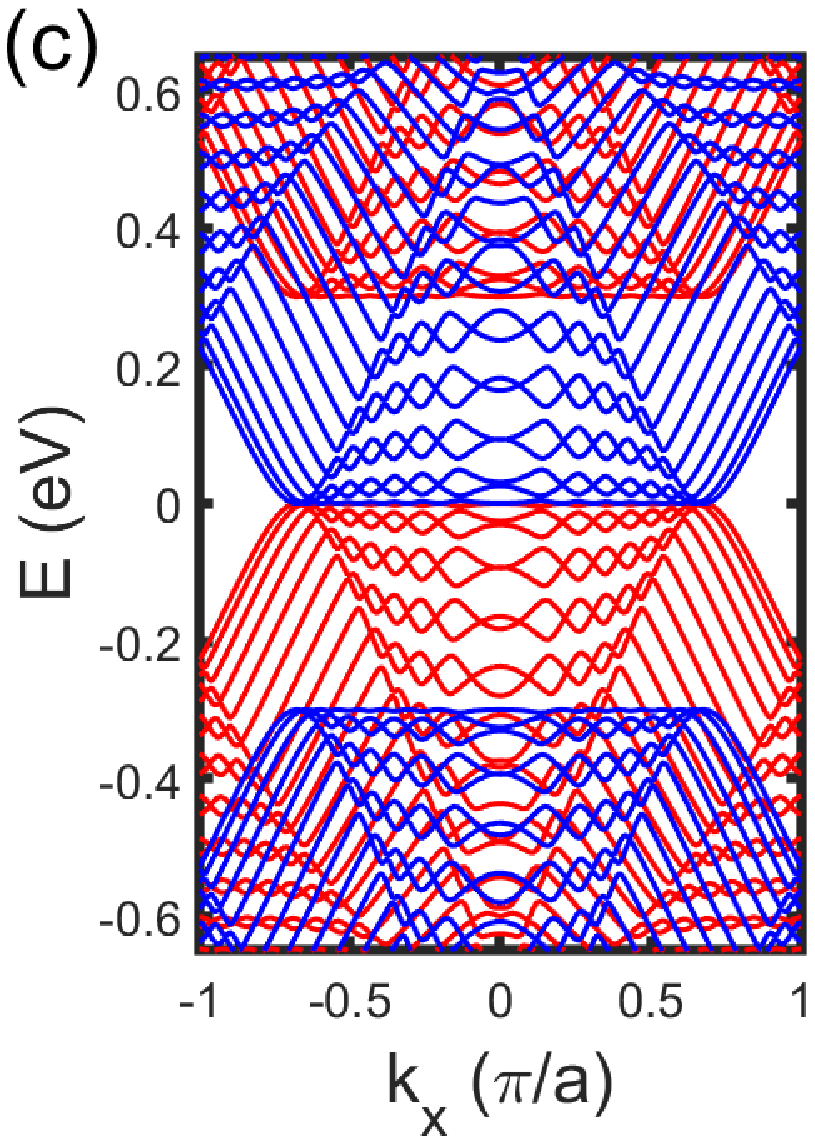}%
    \includegraphics[scale=0.38,trim=0 0 0 0,clip]{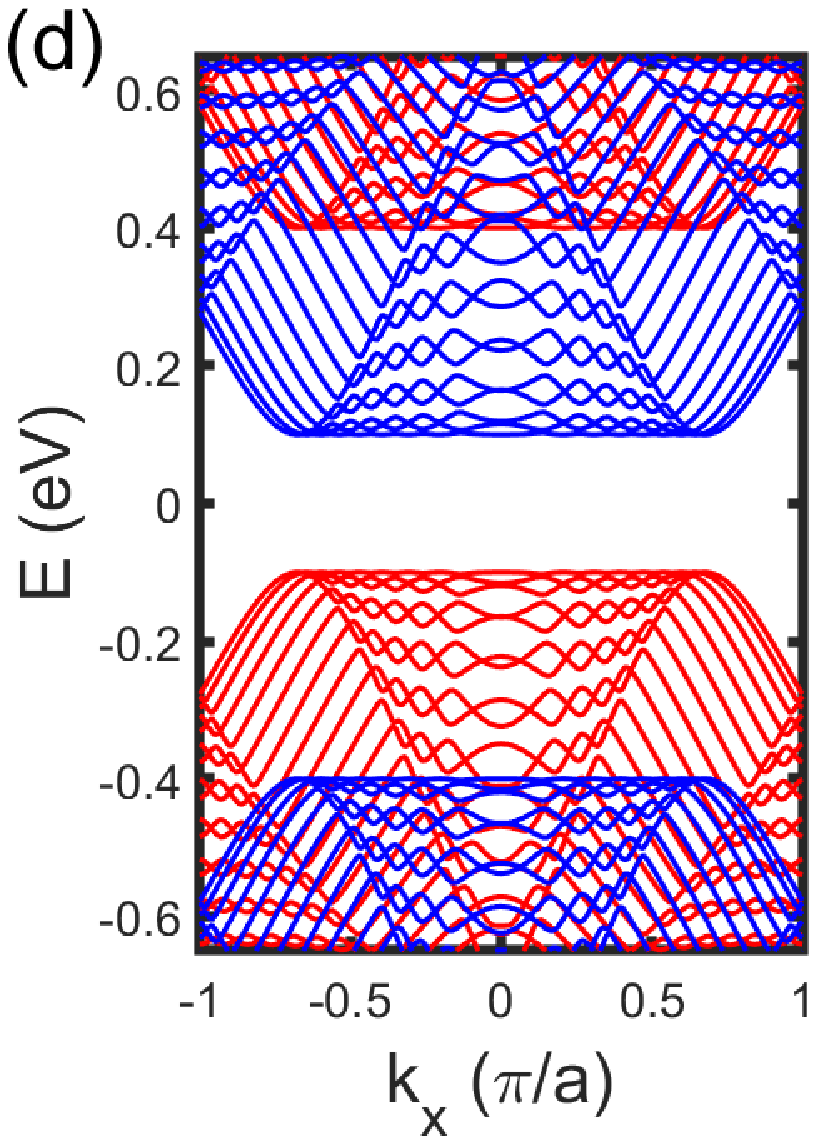}%
	\caption{Spin-dependent energy bands of ATSNR for different sublattice potentials (a) $E_\xi=0$, (b)$E_\xi=0.10$ eV, (c) $E_\xi=0.15$ eV, (d) $E_\xi=0.25$ eV. Red (blue) curves represent spin-up (spin-down) energy bands. The exchange field is $M=0.15$ eV, other parameters are the same as those in Fig.~\ref{fig:band}.}
	\label{fig:spinband2}
\end{figure}

Next we discuss the effect of the sublattice potential on the energy bands for the fixed exchange field $M=0.15$ eV.
In Fig. \ref{fig:spinband2}(a), there exists no band gap and spin degeneracy of the energy band is broken with spin splitting of $0.3$ eV
because of the exchange field. When an electric field is applied on the central region, spin-up and spin-down energy bands are broken and there
exist spin-dependent band gaps. However, the total band structures have no band gap due to the overlapping of the minimum spin-down conduction band and the maximum spin-up valence band. When $E_\xi=0.15$ eV, the energy bands become identical with those given in Fig. \ref{fig:spinband1}(c).
With $E_\xi$ increasing to 0.25 eV, the spin-dependent band gap is 0.5 eV, while the spin-independent band gap is 0.2 eV.
Obviously, exchange fields can break spin degeneracy and induce spin splitting. When $E_\xi>M$, we can see that there exists a band gap.
\begin{figure}[t]
	\centering
	\includegraphics[scale=0.38,trim=0 0 0 0,clip]{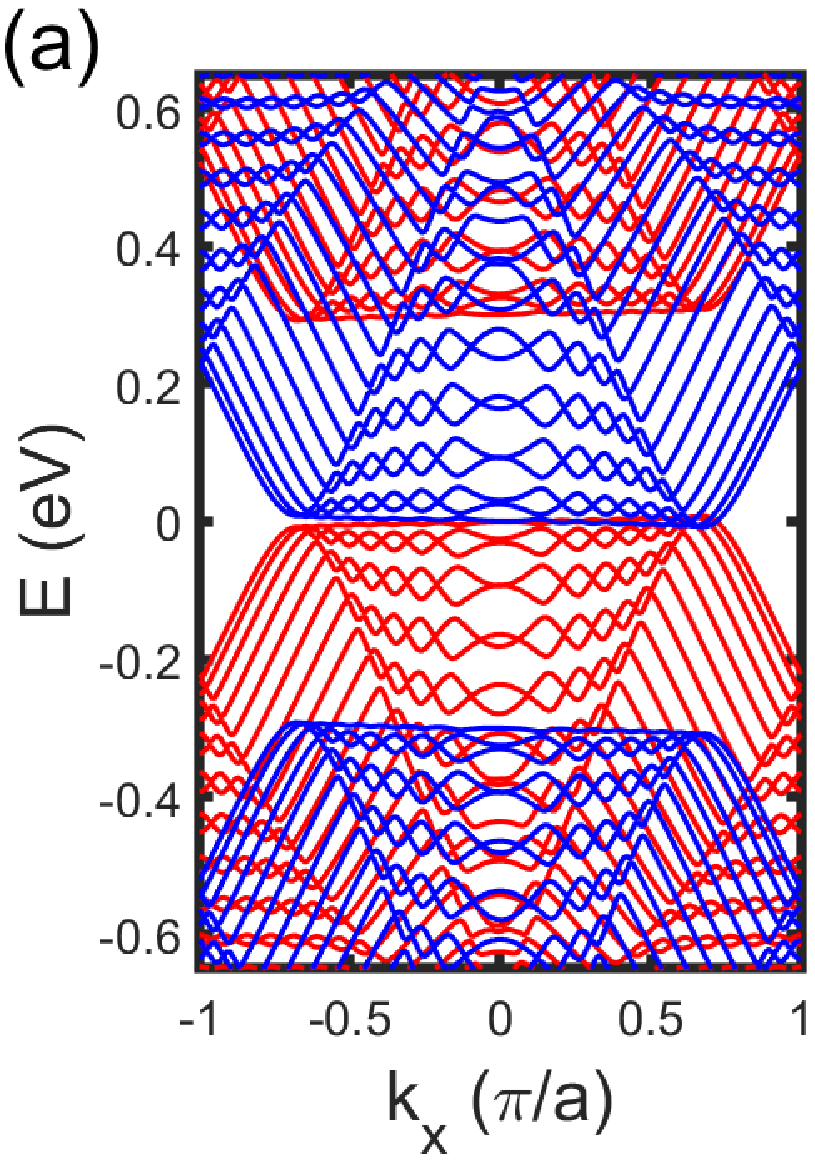}%
	\includegraphics[scale=0.38,trim=0 0 0 0,clip]{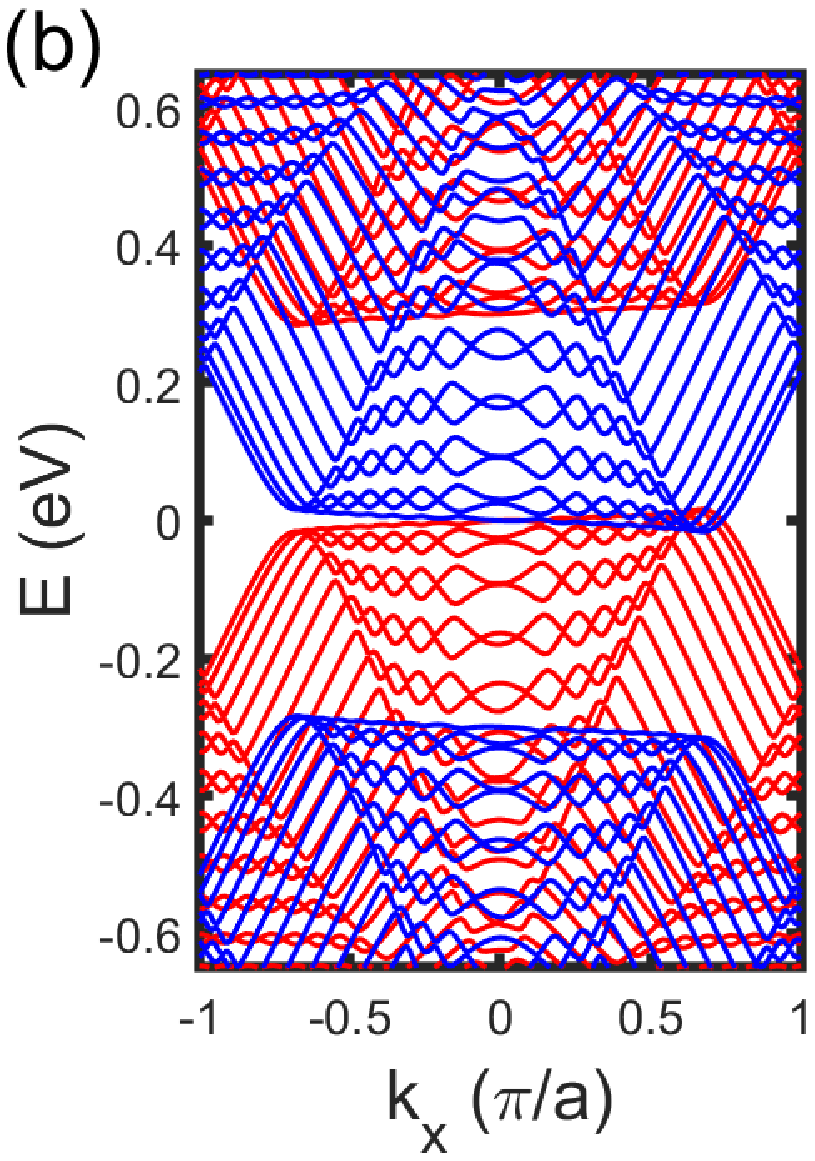}

	\includegraphics[scale=0.38,trim=0 0 0 0,clip]{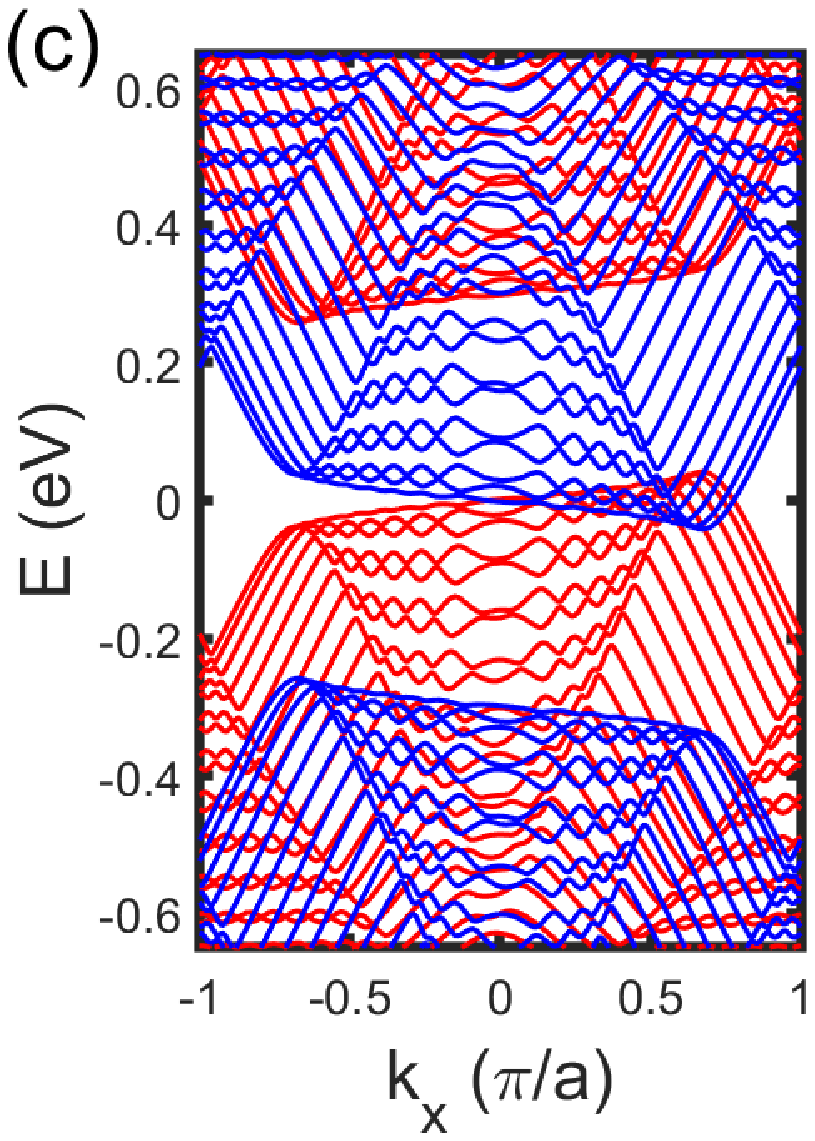}%
	\includegraphics[scale=0.38,trim=0 0 0 0,clip]{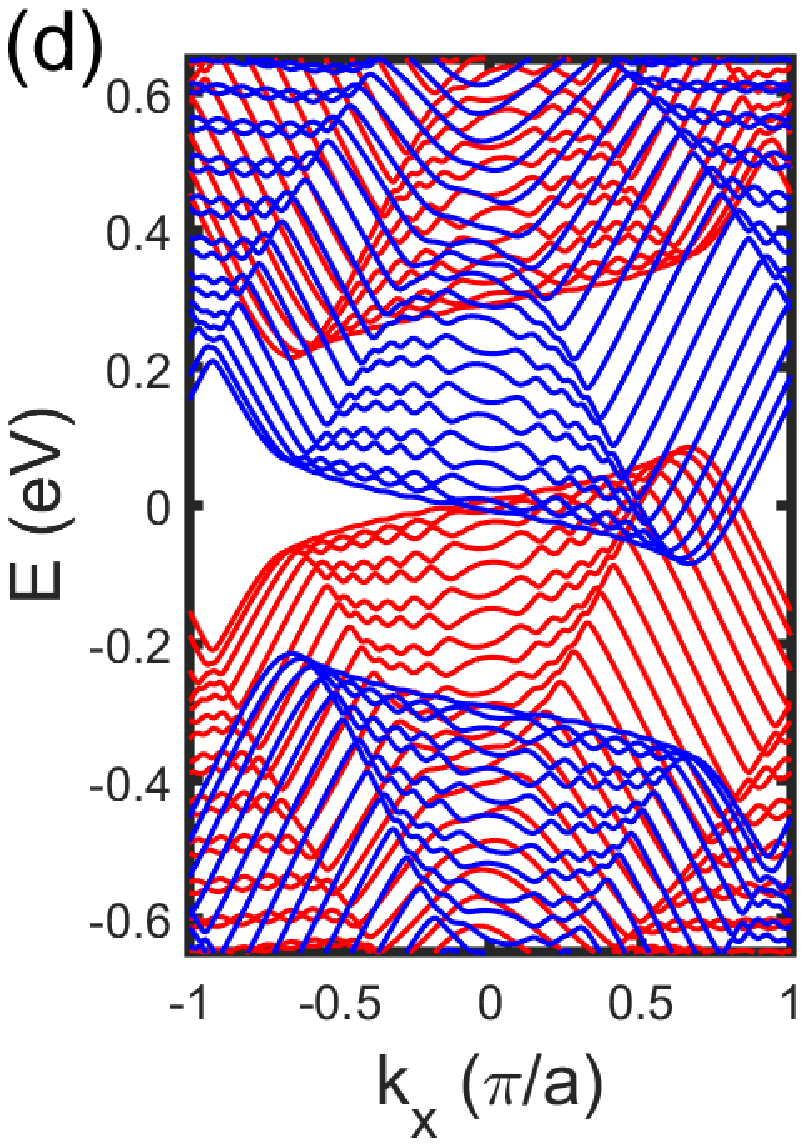}
	\caption{Spin-dependent energy bands of ATSNR for different strengths of SOC with (a) $t_{so}=0.01$eV, (b) $t_{so}=0.02 $eV, (c) $t_{so}=0.05$eV, and (d) $t_{so}=0.1$eV. The sublattice potential and the exchange field are $E_{\xi}=M=0.15$eV. Other parameters are the same as those in Fig.~\ref{fig:band}.}
	\label{fig:SOCband1}
\end{figure}

\subsection{Effect of the spin-orbit coupling}
\begin{figure}[t]
	\centering
	\includegraphics[scale=0.38,trim=0 0 0 0,clip]{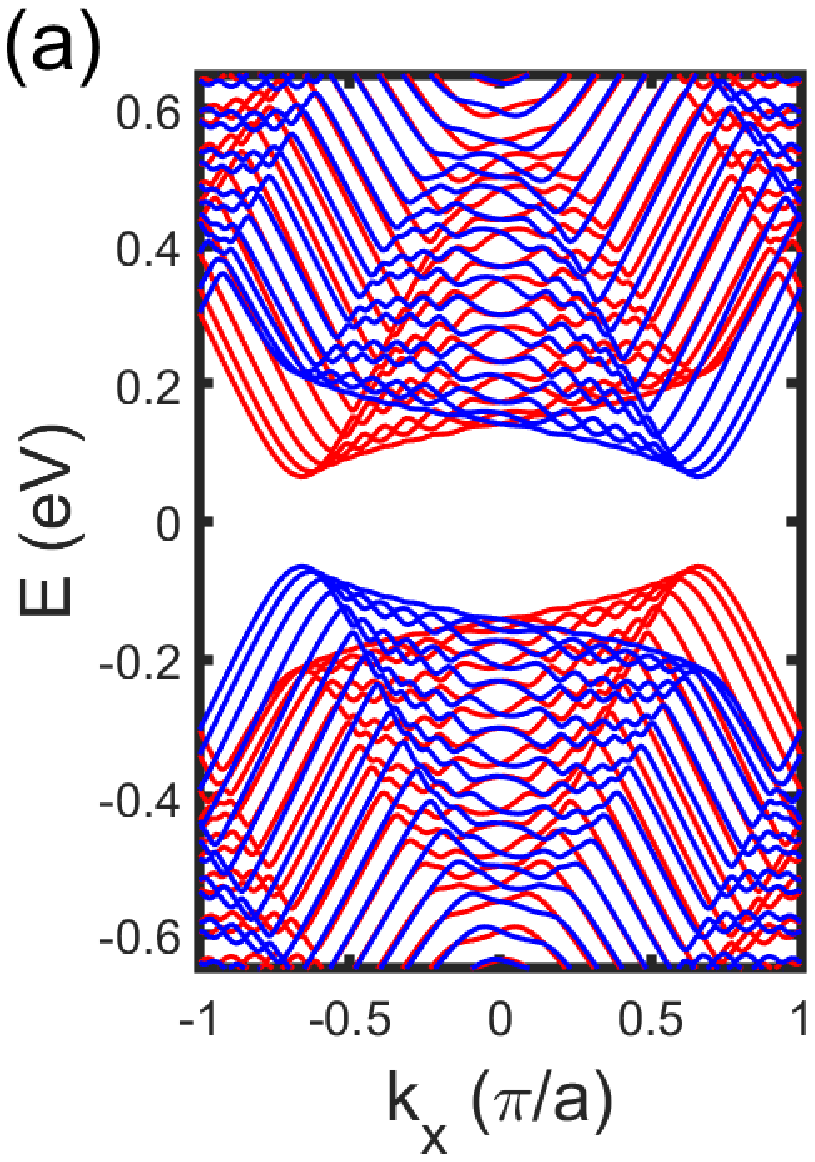}
	\includegraphics[scale=0.38,trim=0 0 0 0,clip]{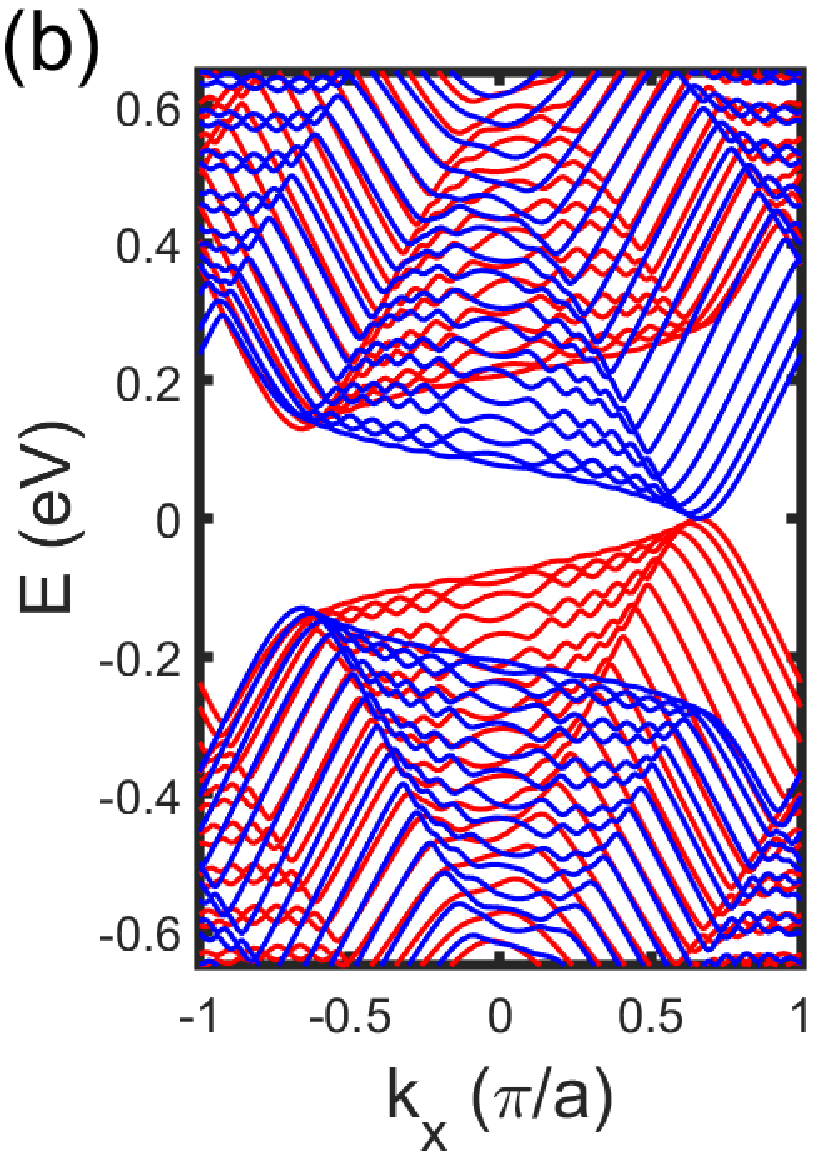}

	\includegraphics[scale=0.38,trim=0 0 0 0,clip]{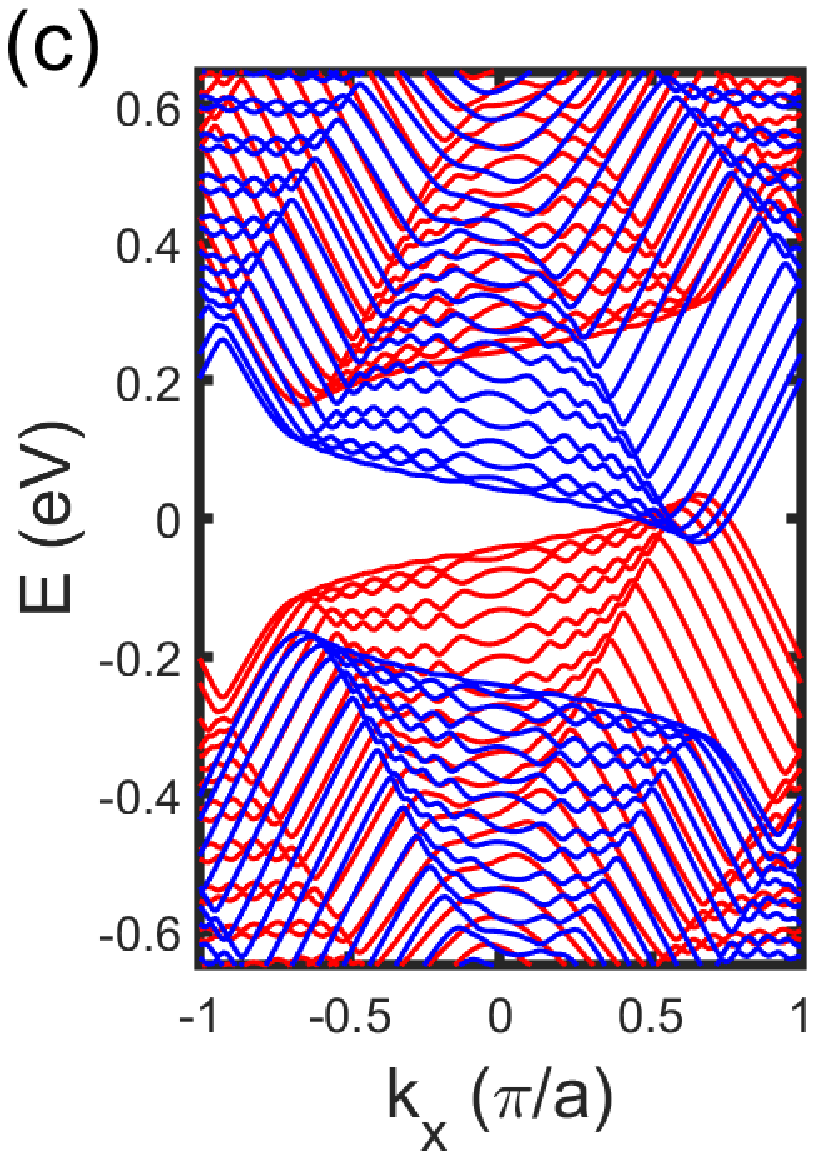}
	\includegraphics[scale=0.38,trim=0 0 0 0,clip]{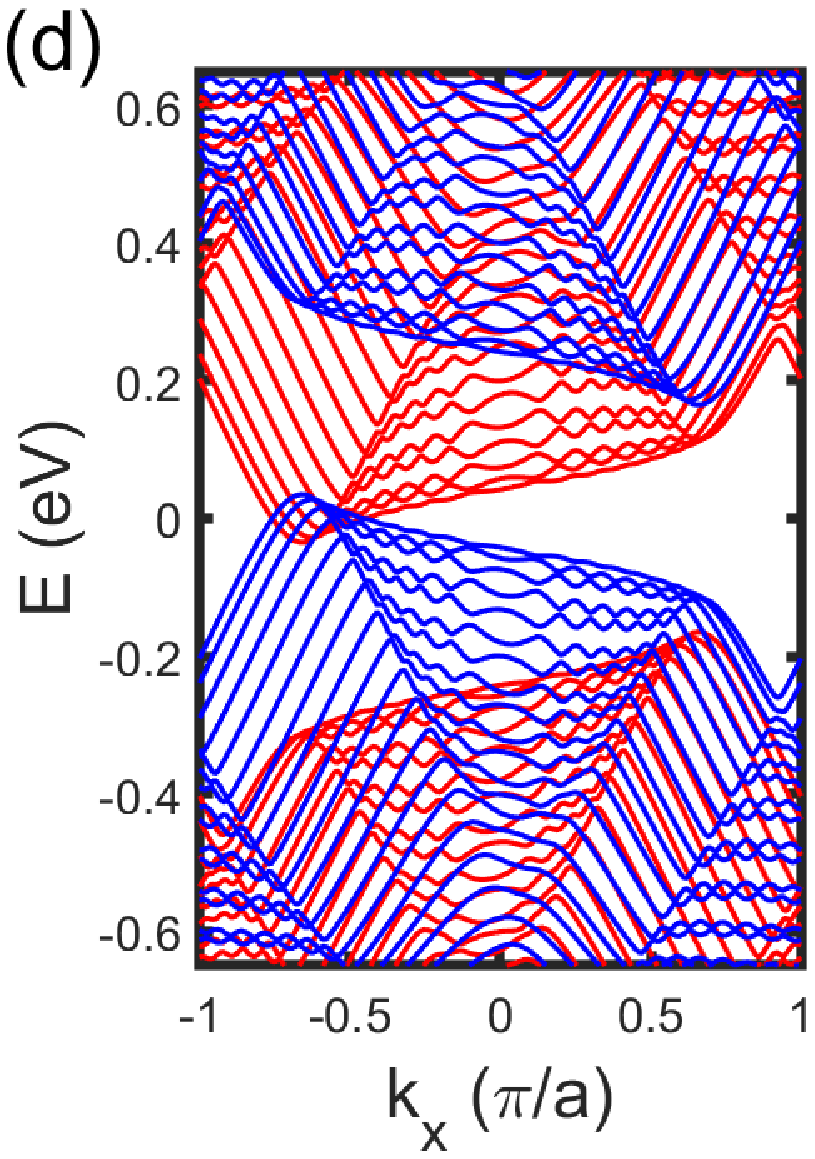}
	\caption{Spin-dependent energy bands of ATSNR for different exchange fields (a) $M=0.05$eV, (b) $M=0.07$eV, (c) $M=0.10$eV, and (d) $M=-0.10$eV. The spin-orbit coupling is $t_{so}=0.1$eV, the sublattice potential is $E_\xi=0.15$eV. Other parameters are the same as those in Fig.~\ref{fig:band}.}
	\label{fig:SOCband2}
\end{figure}
The low-buckled structure of silicene systems induces a relatively large spin-orbit coupling associating with the next-nearest-neighbor hopping term~\cite{Verri,Liu}. Since the tetragonal silicene also has a low-buckled structure, it is possible for the next-nearest-neighbor hopping term to induce
a spin-orbit coupling. So far, there has few research work in literature reporting the effect of the spin-orbit coupling in the tetragonal silicene. In this paper, we first propose to include the contribution of the spin-orbit coupling by adopting phenomenological next-nearest-neighbor hopping
parameters, which need to be verified by future experimental measures or calculating results obtained from DFT method.

Thus an extra Hamiltonian arising from the next-nearest-neighbor hopping terms can be written as~\cite{SOC1}
\begin{eqnarray}\label{eq:parameter}
\begin{aligned}
	H_{SOC}=it_{so}\sum_{\langle\langle i,j \rangle\rangle,\alpha\beta}{\nu}_{ij}{ c_{i\alpha}^\dag\sigma}^z_{\alpha\beta} c_{j\beta}
 \\+ it_R\sum_{\langle\langle i,j\rangle\rangle,\alpha\beta}{\mu}_ic_{i\alpha}^\dag(\vec{\sigma}\times \mathbf{d}_{ij})^z_{\alpha\beta} c_{j\beta},
\end{aligned}
\end{eqnarray}
here the first term represents the intrinsic SOC, and the second term refers to Rashba SOC.
The amplitude $t_R$ of Rashba SOC is relatively smaller. It is find that Rashba term does not cause an effective spin splitting, which is neglected in the following calculations. As shown in Fig. \ref{fig:lead}(c), the tetragonal silicene has only two kinds of next-nearest-neighbor hopping modes, namely $t_{so1}$ and $t_{so2}$ corresponding to the hopping distances $L_4$ and $L_3$, respectively. Since the two distances satisfy the relation $L_4\approx 1.29 L_3$,
we assume that two SOC hopping terms have a fixed proportional relationship, namely $t_{so2}=\eta t_{tso1}=t_{so}$ with $\eta=1.3$.

Spin-dependent energy bands of ATSNR for different strengths of SOC are plotted as a function of the wave vector $k_x$, as shown in Fig.~\ref{fig:SOCband1}.
It can be seen that SOC can result in tilted subbands. What's more, spin-up energy bands move towards upper right portion, while spin-down energy bands move towards upper left portion. With increasing of the strength of SOC, the tilted angle gradually increases. For example, when $t_{so}=0.1$eV, there exist obvious overlapping subbands in the range of $k_x>0$. While a large band gap is induced by SOC in the range of $k_x<0$.

Next we study the combined effect of the exchange field, SOC and the sublattice potential on the energy bands. When $E_\xi=0.15$eV and $M=0.05$eV, there exists a band gap and spin splitting in the absence of SOC, which associates with the energy bands given in Fig.~\ref{fig:spinband1}(b). In Fig.~\ref{fig:SOCband2}, considering the effect of SOC with $t_{so}=0.1$eV, spin-up and spin-down subbands tilted towards opposite directions. With the exchange field increasing to 0.07eV, the spin-down subbands move downwards, while the spin-up subbands move upwards. Thus the spin-down conduction band and the spin-up valence band begin to touch each other near $k_x\approx 0.6\pi/a$. Interestingly,
the tilted angles of the spin-dependent subbands keep unchanged when SOC is fixed to be 0.1eV. When $M$ increases to 0.1eV, the spin-dependent subbands overlap and the bang gap disappears. When the exchange field is changed to the opposite direction, the overlapping subbands locate at the left region with negative values for the wave vector $k_x$, as shown in Fig.~\ref{fig:SOCband2}(d)

\subsection{Spin-dependent conductance and spin polarization}
\begin{figure}[t]
	\centering
	\includegraphics[scale=0.30,trim=0 0 0 0,clip]{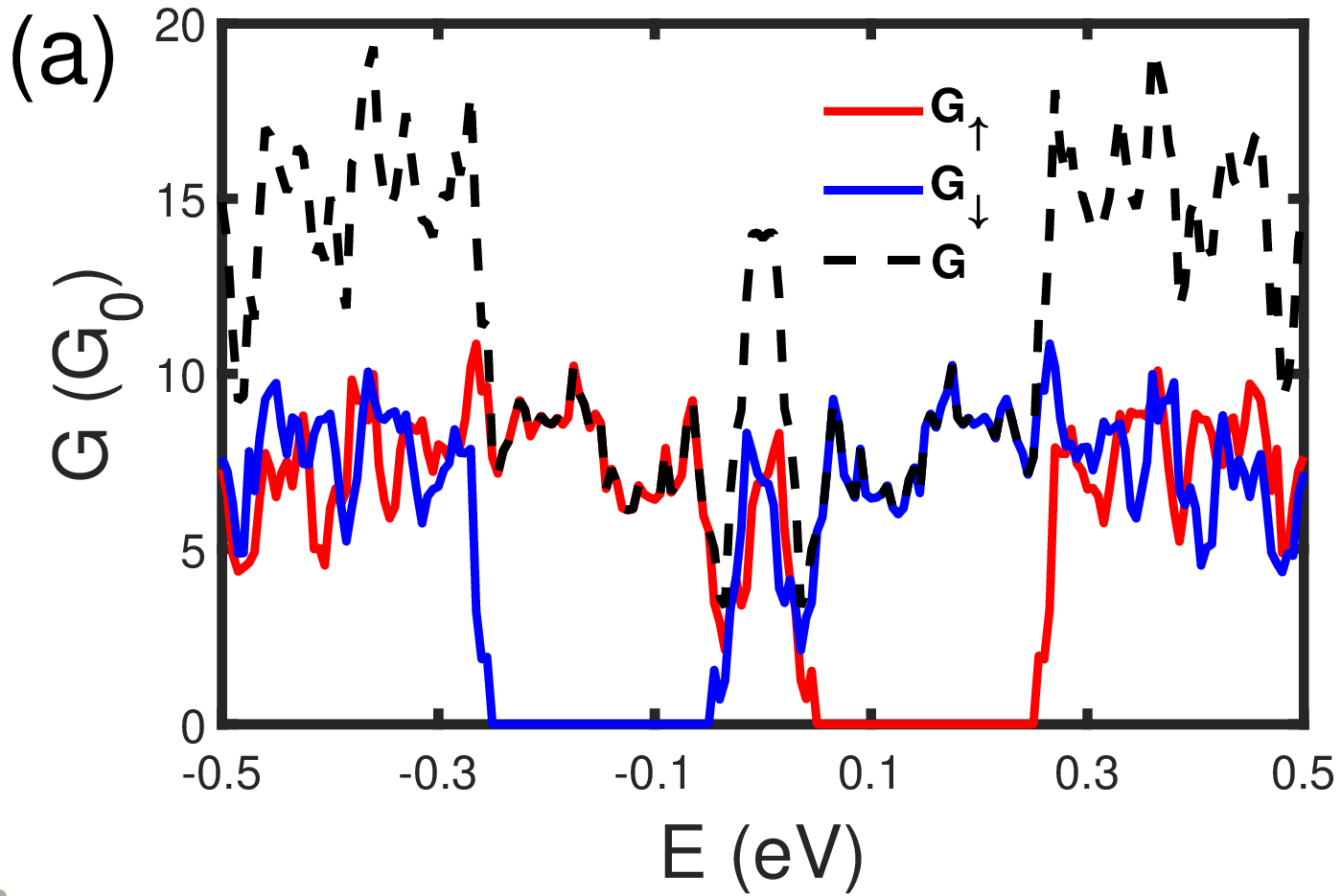}%
    \includegraphics[scale=0.30,trim=0 0 0 0,clip]{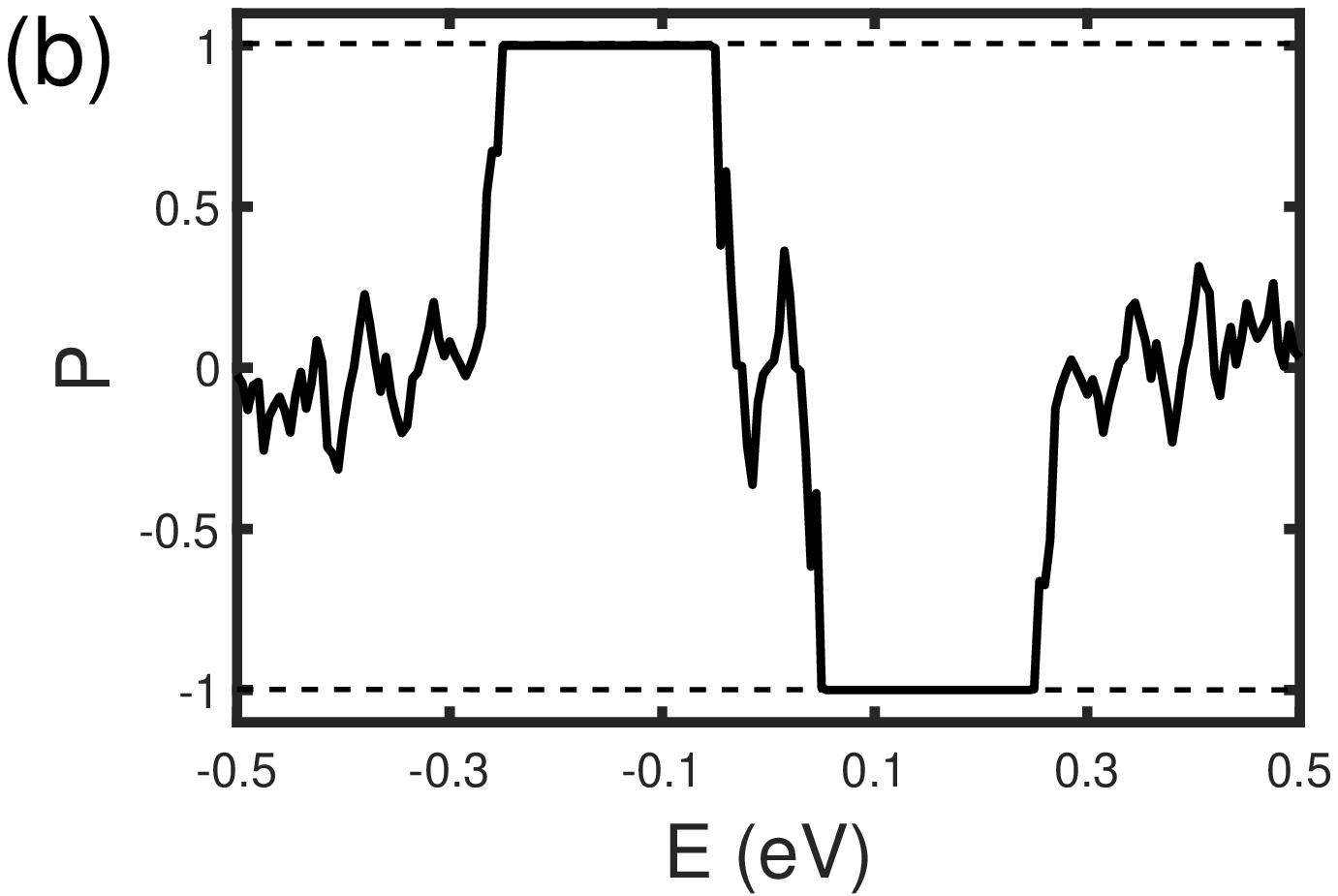}%

    \includegraphics[scale=0.30,trim=0 0 0 0,clip]{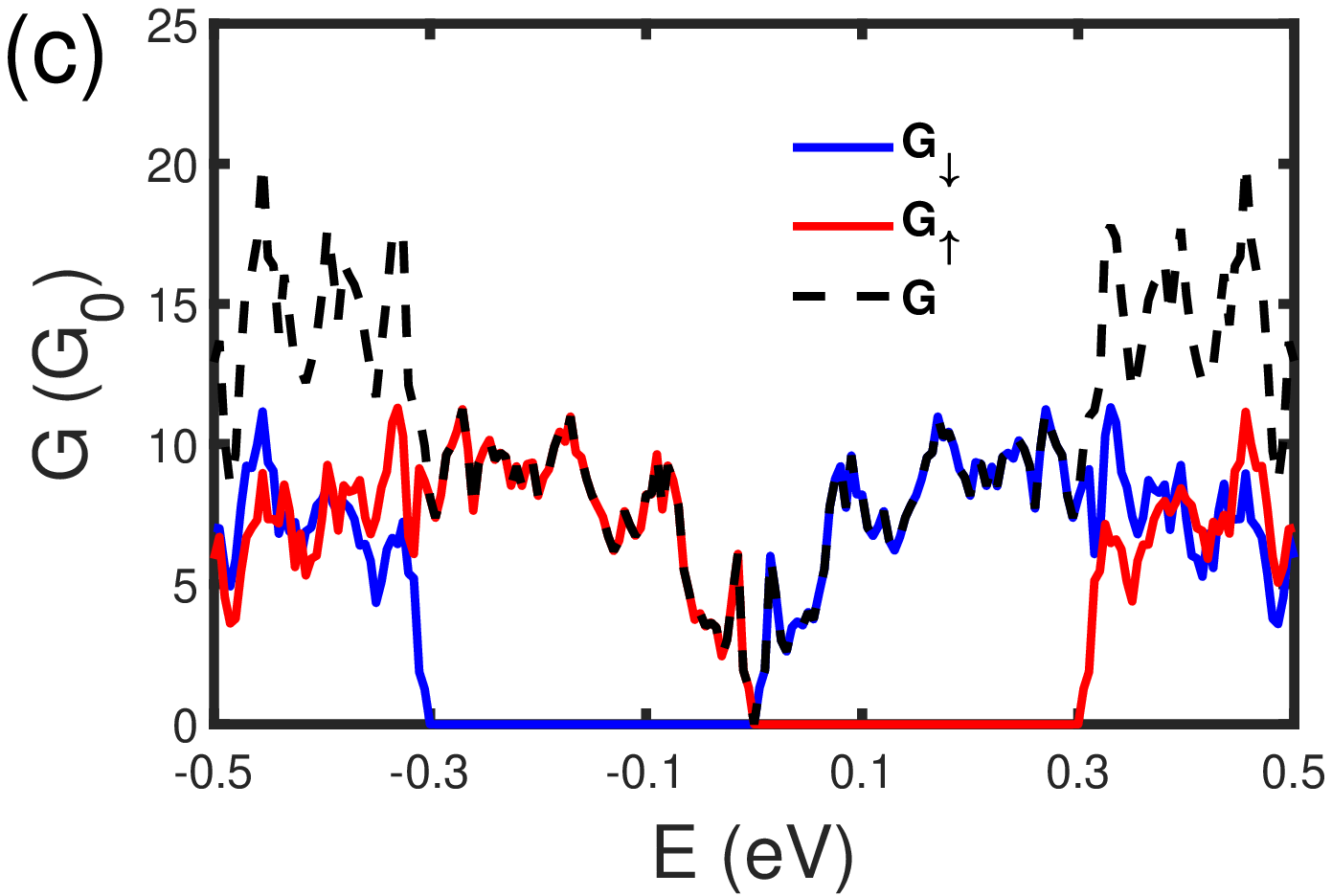}%
    \includegraphics[scale=0.30,trim=0 0 0 0,clip]{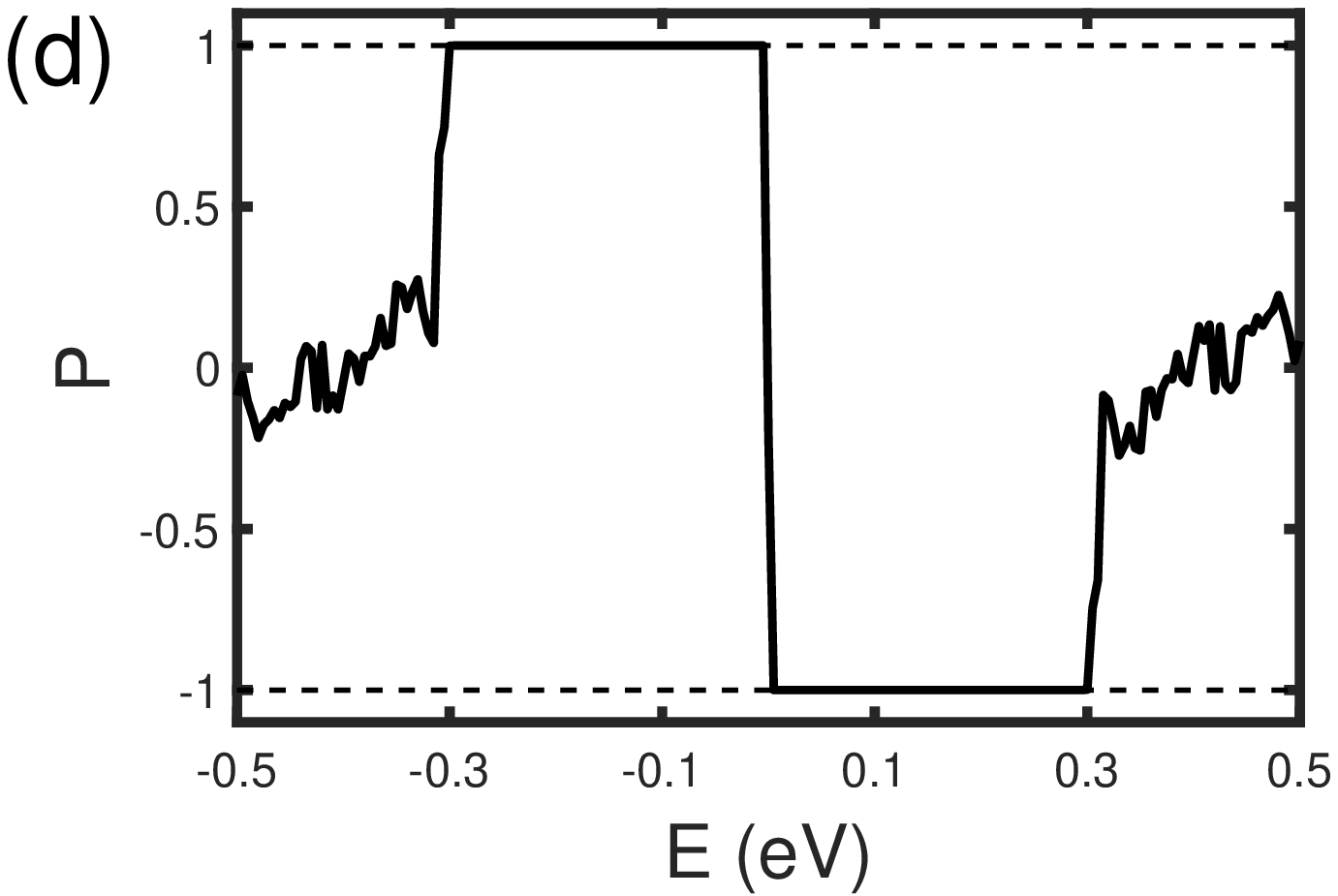}%

    \includegraphics[scale=0.30,trim=0 0 0 0,clip]{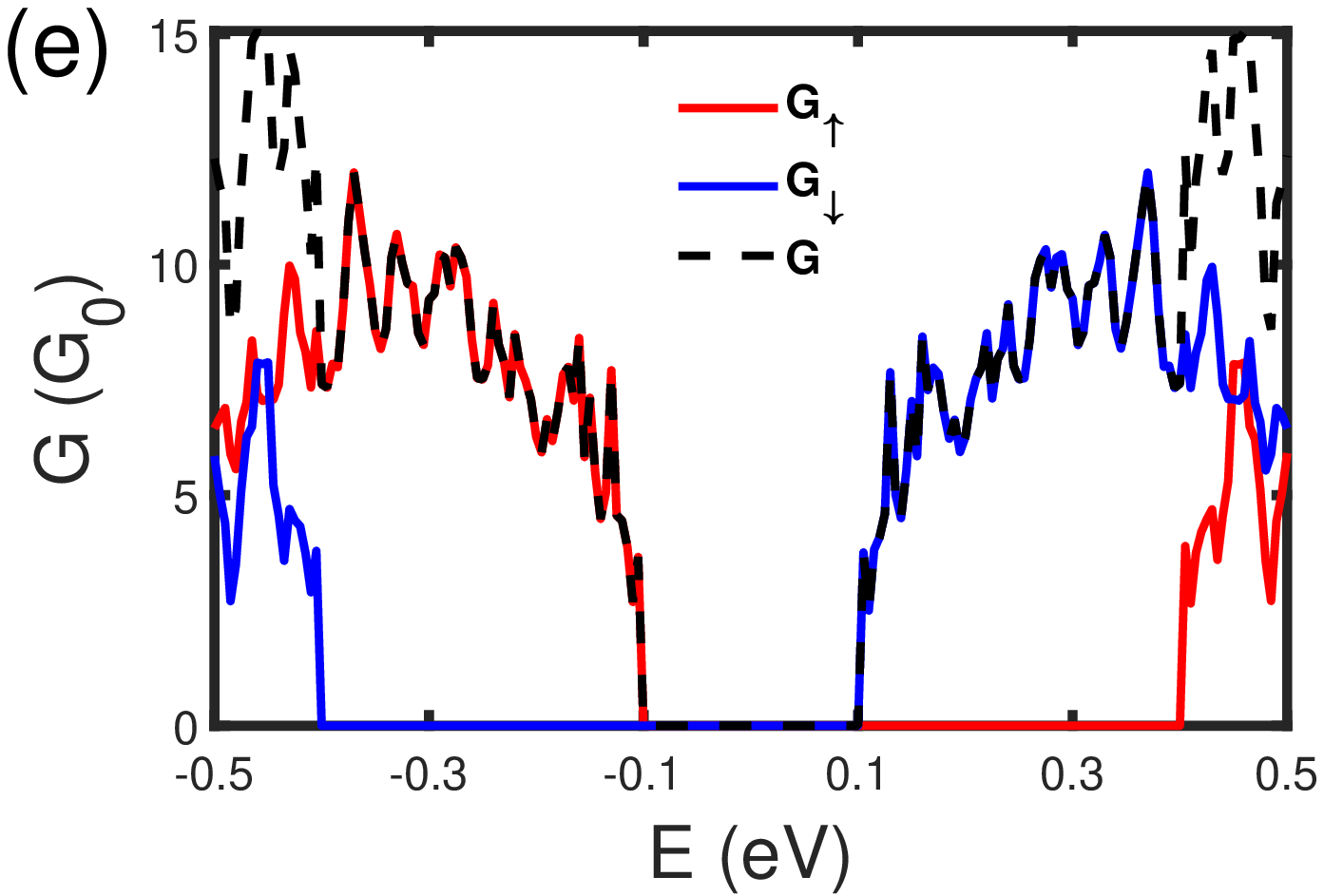}%
    \includegraphics[scale=0.30,trim=0 0 0 0,clip]{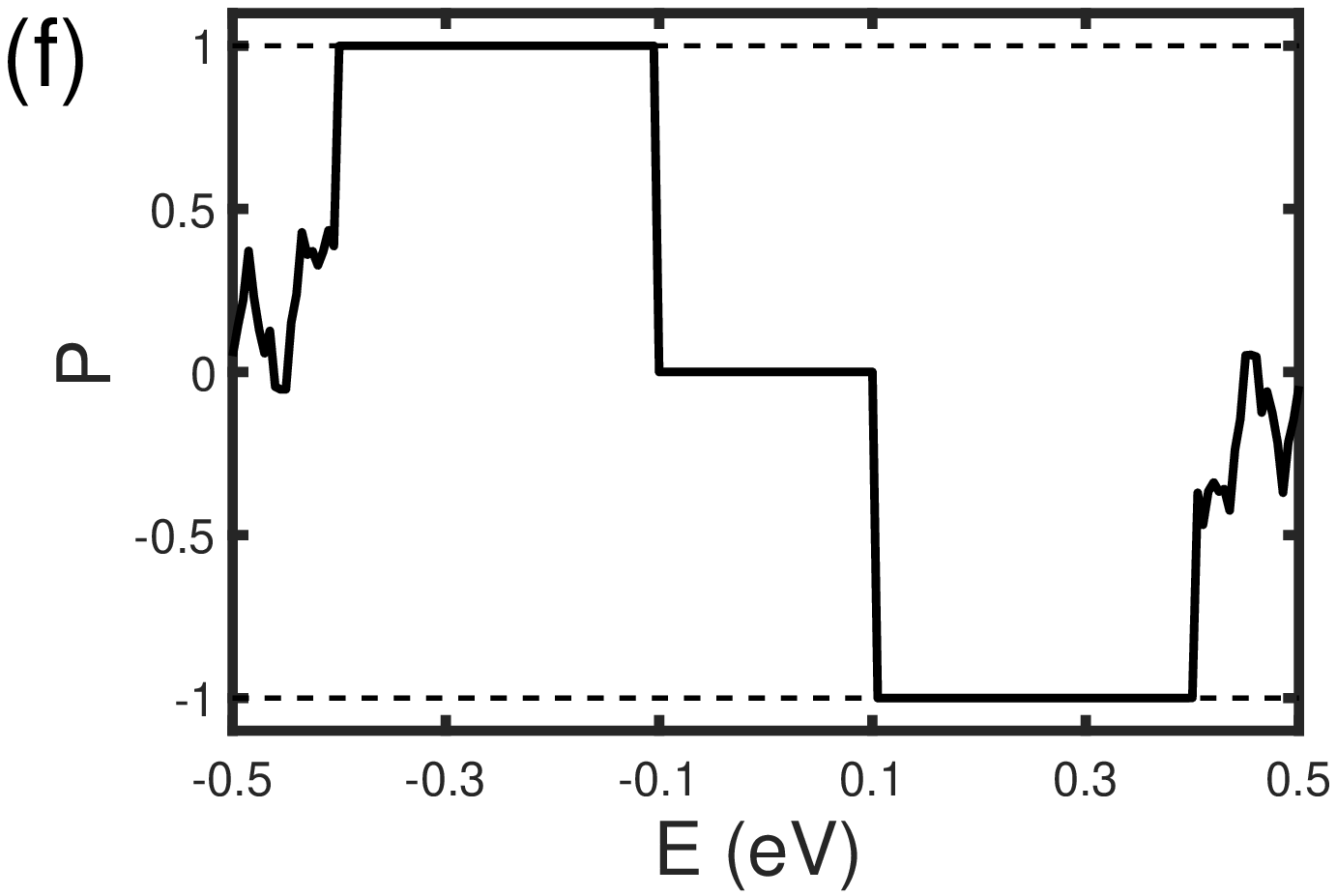}%
	\caption{Spin-dependent conductance and spin polarization are plotted  as a function of Fermi energy for different sublattice potentials (a), (b) $E_\xi=0.1$ eV; (c), (d) $E_\xi=0.15$ eV; (e), (f) $E_\xi=0.25$ eV. Red (blue) curves represent spin-up (spin-down) conductance, and black dash curve represents the total conductance. The exchanged field is $M=0.15$ eV, other parameters are the same as those in Fig.~\ref{fig:band}.}
	\label{fig:GP}
\end{figure}
We further study the combined effect of the exchange field and the electric field on the spin-dependent transport property.
Note that we first neglect the effect of SOC.
The spin-dependent conductance can be calculated in terms of Eq.~(\ref{eq:Cond}). The spin polarization is defined as
 \begin{eqnarray}
	P=\frac{G_{\uparrow}-G_{\downarrow}}{G_{\uparrow}+G_{\downarrow}}.
\label{eq:parameter3}
\end{eqnarray}
When $E_\xi=0.1$ eV, as shown in Fig.~\ref{fig:GP}(a), the spin-up (spin-down) conductance $G_\uparrow$ ($G_\downarrow$) oscillates with increasing of Fermi energy.  The maximum value of $G_\uparrow$ ($G_\downarrow$) is about $11G_0$. Interestingly, the spin-up conductance is zero in the energy range of $0.05~\mathrm{eV}\leq E\leq 0.25~\mathrm{eV}$, while the spin-down conductance is zero in the energy range of $-0.25~\mathrm{eV}\leq E\leq -0.05~\mathrm{eV}$.
The zero-conductance behaviors associate with energy bands given in Fig.~\ref{fig:spinband2}(b). Correspondingly, the spin polarization becomes P=1 or -1 [see Fig.~\ref{fig:GP}(b)] in the two zero-conductance ranges because the transmitted electrons are completely spin polarized. The total conductance has no zero value due to the overlapping of spin-down conduction bands and spin-up valence bands. Obviously, separation of spin-dependent conductances arises from the effect of the exchange field, while zero-conductance behaviors exhibit spin-dependent band gaps induced by the electric field.

When $E_\xi=0.15$ eV, $G_\uparrow$ has zero value in the energy range $0\leq E\leq 0.3~\mathrm{eV}$, and $G_\downarrow$ is zero in the energy range $-0.3~\mathrm{eV}\leq E\leq 0$. Similarly, in the two zero-conductance ranges, the electrons are still completely spin polarized, which give an abrupt transition for the spin polarization changing from P=1 to P=-1 at $E=0$. The flatted band at $E=0$ results in a zero value of the total conductance.
With the sublattice potential increasing to $E_\xi=0.25$ eV [see Fig.~\ref{fig:GP}(e)], the energy range for the spin-dependent conductance with zero value broadens to 0.5 eV. However, for this case, $G_\uparrow$ and $G_\downarrow$ have zero values in a common energy range with $-0.1~\mathrm{eV}\leq E\leq 0.1~\mathrm{eV}$, which associates with a band gap shown in Fig.~\ref{fig:spinband2}(d) thus resulting in zero total conductance. When the total conductance is zero, we set spin polarization to be zero, namely P=0, which cannot be obtained in terms of the definition given in Eq.~(\ref{eq:parameter3}).

\begin{figure}[t]
	\centering
	\includegraphics[scale=0.33,trim=80 0 0 0,clip]{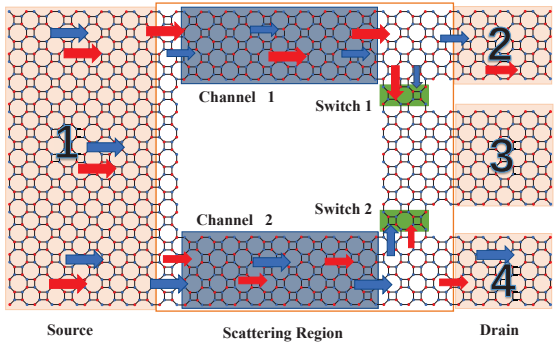}%
	\caption{Schematic of four-terminal two channels spin-based ATSNR. Red and blue arrows indicate spin currents of spin-up and spin-down electrons, respectively. Channel 1 and 2 can achieve a fully spin-polarized current. The green region acts as a switch and inhibits the flow of electrons by applying a high potential barrier $V_s$.}
	\label{fig:device}
\end{figure}

\subsection{Generation of fully spin-polarized current via two-channel device in the absence of SOC}
In terms of above analyses, it is expected to realize a fully spin-polarized current by using two-channel spin-based ATSNR.
Thus, we propose a device configuration of four-terminal and two channels spin-based ATSNR, as shown in Fig.~\ref{fig:device}.
With help of the latest quantum dot technology, we can chisel the central scattering region into two channels, and select two narrow regions [see green parts] in the right buffer layer to exert high potential barriers in order to control the flowing of electrons, which work as switches.
The central two channels work under the combined effect of the exchange field and the sublattice potential.
The drain lead is divided into three terminals connecting with the right buffer layer.
\begin{figure}[t]
	\centering
    \includegraphics[scale=0.40,trim=0 0 0 0,clip]{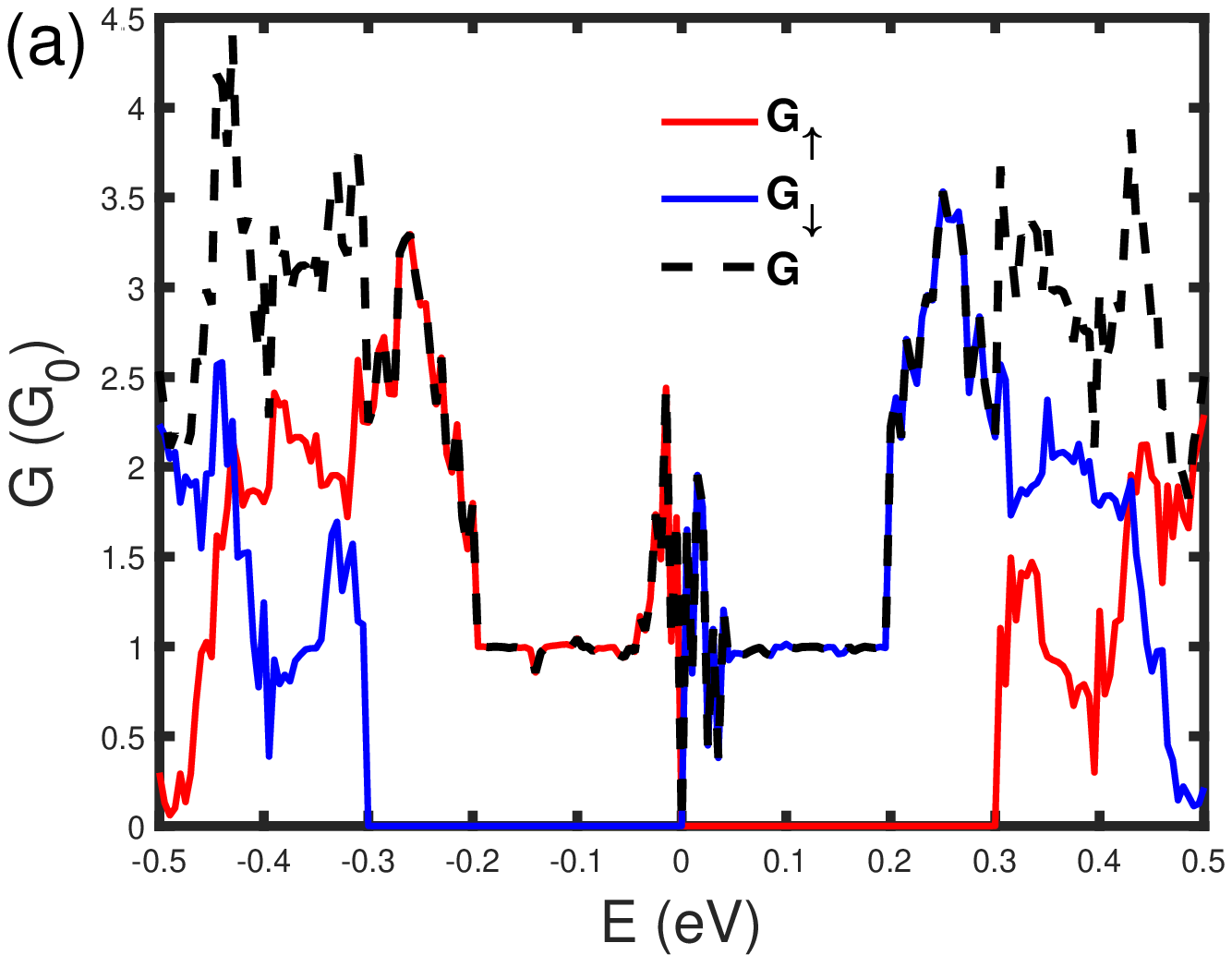}
    \includegraphics[scale=0.40,trim=0 0 0 0,clip]{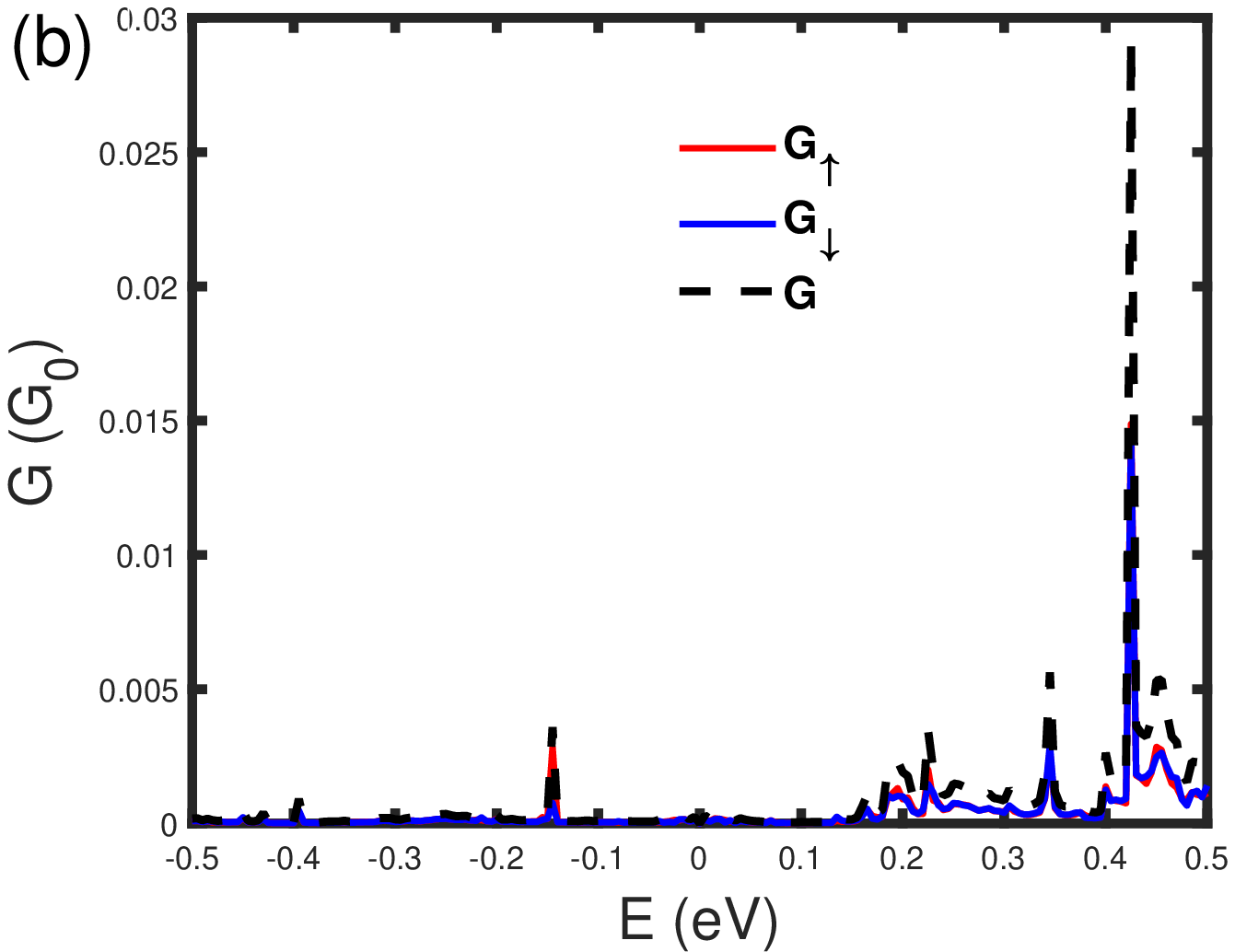}
    \includegraphics[scale=0.40,trim=0 0 0 0,clip]{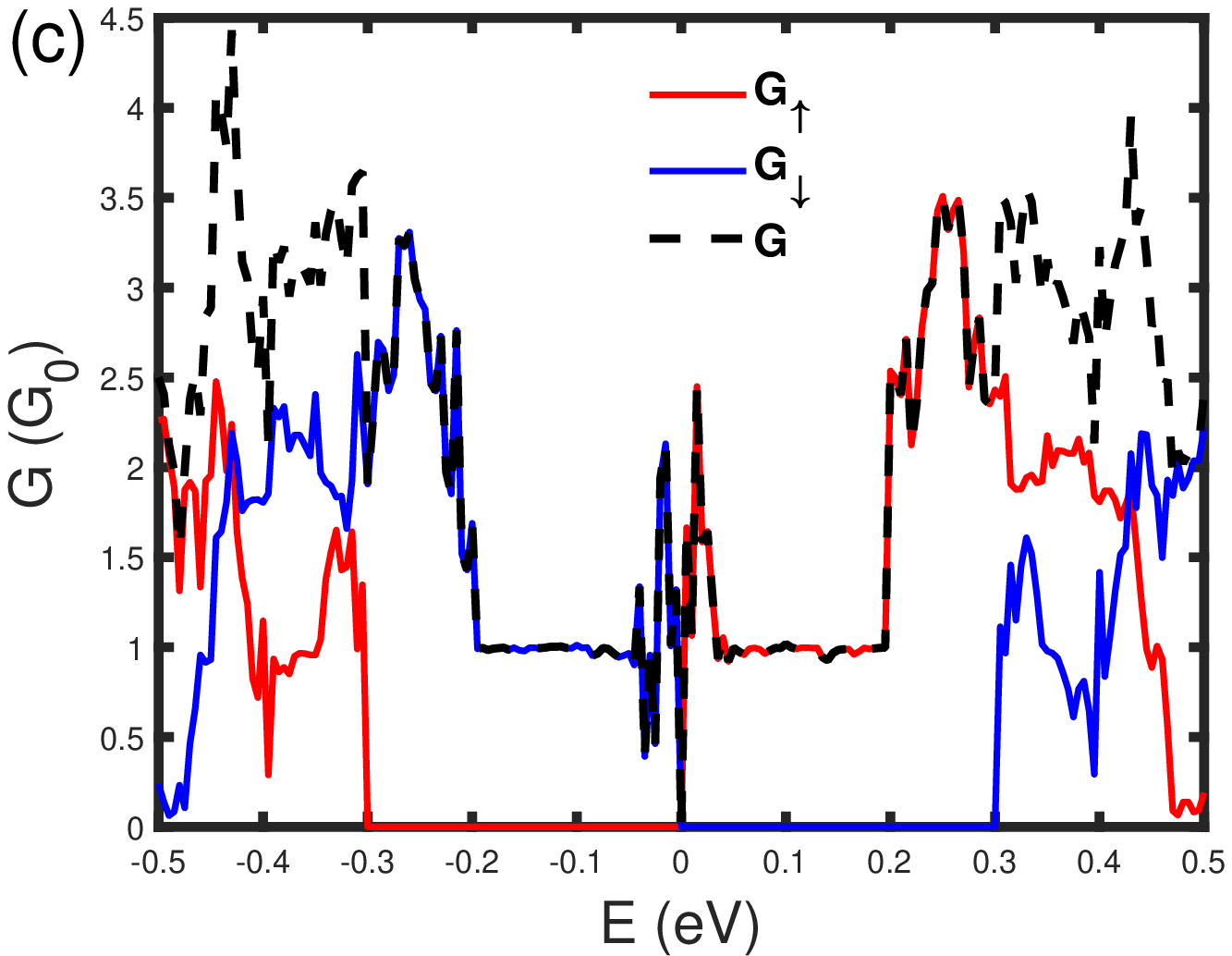}
	\caption{Spin-dependent conductances of two-channel ATSNR for (a) terminal 2, (b) terminal 3 and (c) terminal 4 are plotted  as a function of Fermi energy. Gate voltages of two switches are set to be $V_{1s} = V_{2s} = 4$ eV. Other parameters are given by $E_{\xi 1}=M_1=0.15$ eV in Channel 1 and $E_{\xi 2} = M_2 = -0.15$eV in Channel 2.}
	\label{fig:twoG44}
\end{figure}

We first study the effect of switches on the spin-dependent current for the two-channel ATSNR. It is expected to control the flowing of spin current
by utilizing two switches [see green regions]. Gate voltages of the two switches are set to be $V_{1s} = V_{2s} = 4$ eV. The two channels have opposite amplitudes of exchange fields and sublattice potentials, namely $M_1=E_{\xi 1}=0.15$eV and $M_2=E_{\xi 2}=-0.15$eV.
The spin-dependent conductances of three terminals of the two-channel ATSNR are plotted  as a function of Fermi energy, as shown in Fig.~\ref{fig:twoG44}.
Due to high potential barriers of the two switches, namely $V_{1s} = V_{2s} = 4$ eV, the conductance of terminal 3 [see Fig.~\ref{fig:twoG44}(b)] is much smaller than those of terminal 2 and 4. The maximum conductance of terminal 3 is about $0.03 G_0$, while the maximum conductance of terminal 2 and 4 is about $4.41 G_0$, which is 147 times the conductance of terminal 3. This means that the two switches basically block the electrons flowing from the central scattering region into terminal 3. Accordingly, the spin current of terminal 2 (terminal 4) is basically modulated by the parameters of channel 1 (channel 2).
\begin{figure}[t]
	\centering
    \includegraphics[scale=0.4,trim=0 0 0 0,clip]{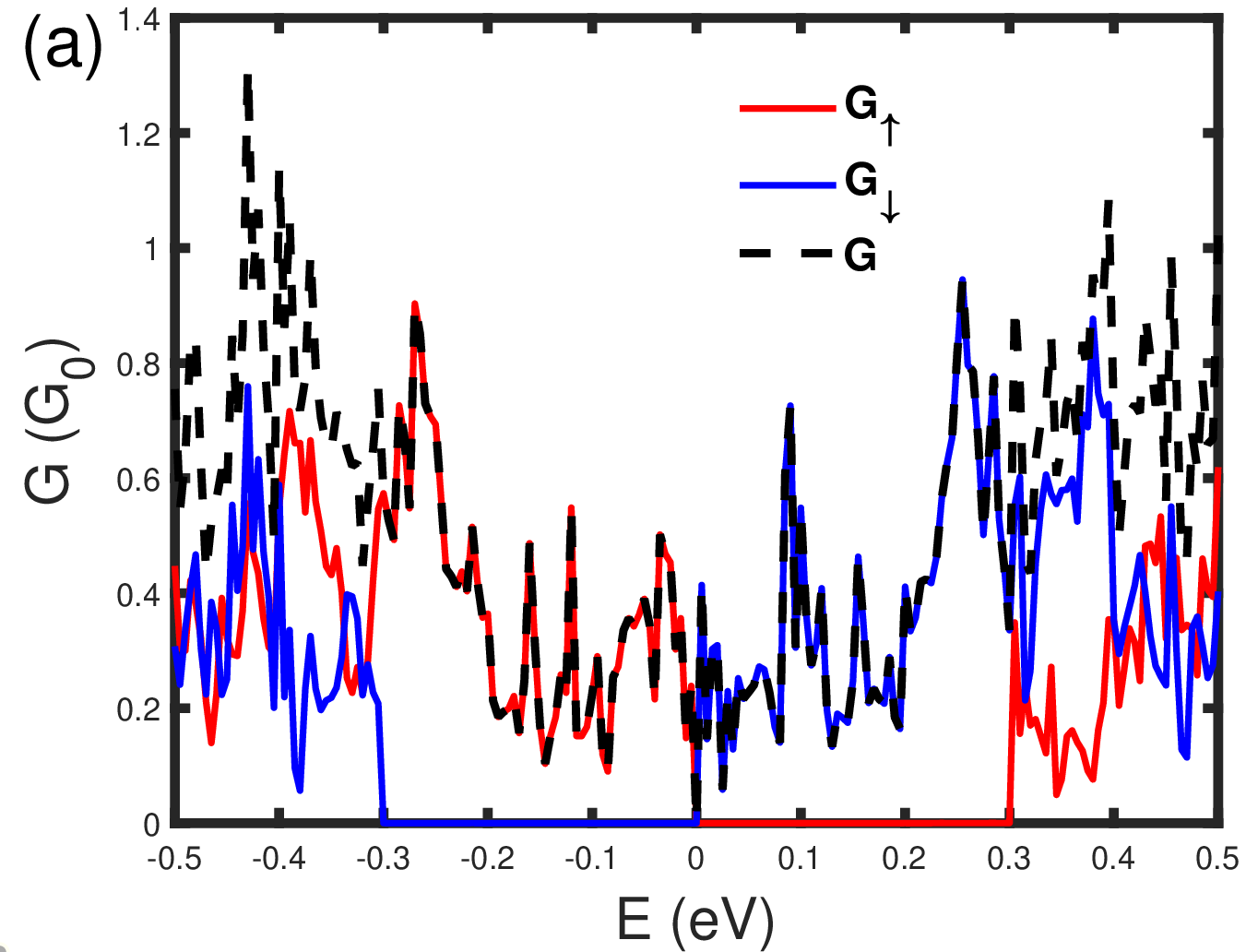}
    \includegraphics[scale=0.4,trim=0 0 0 0,clip]{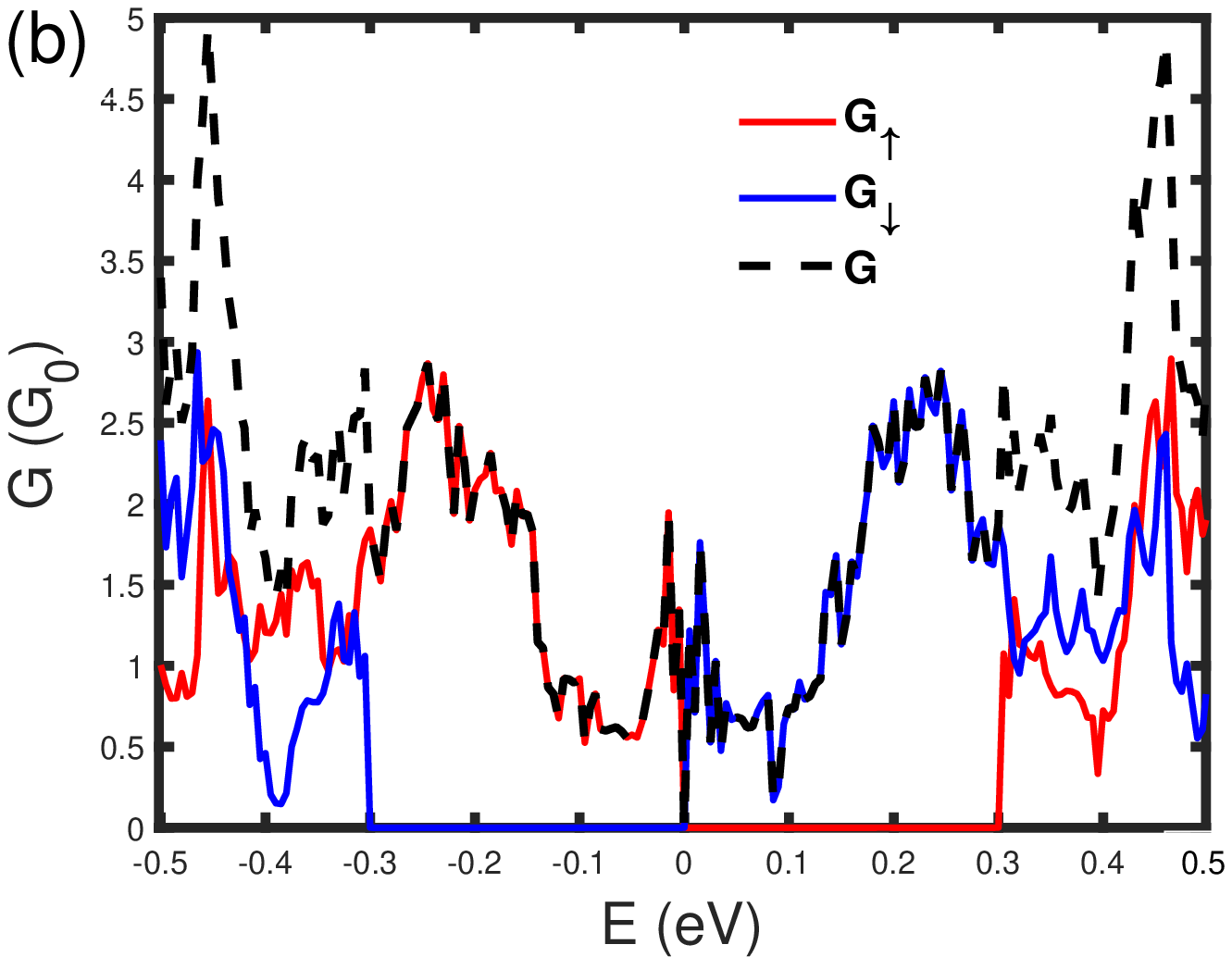}
    \includegraphics[scale=0.4,trim=0 0 0 0,clip]{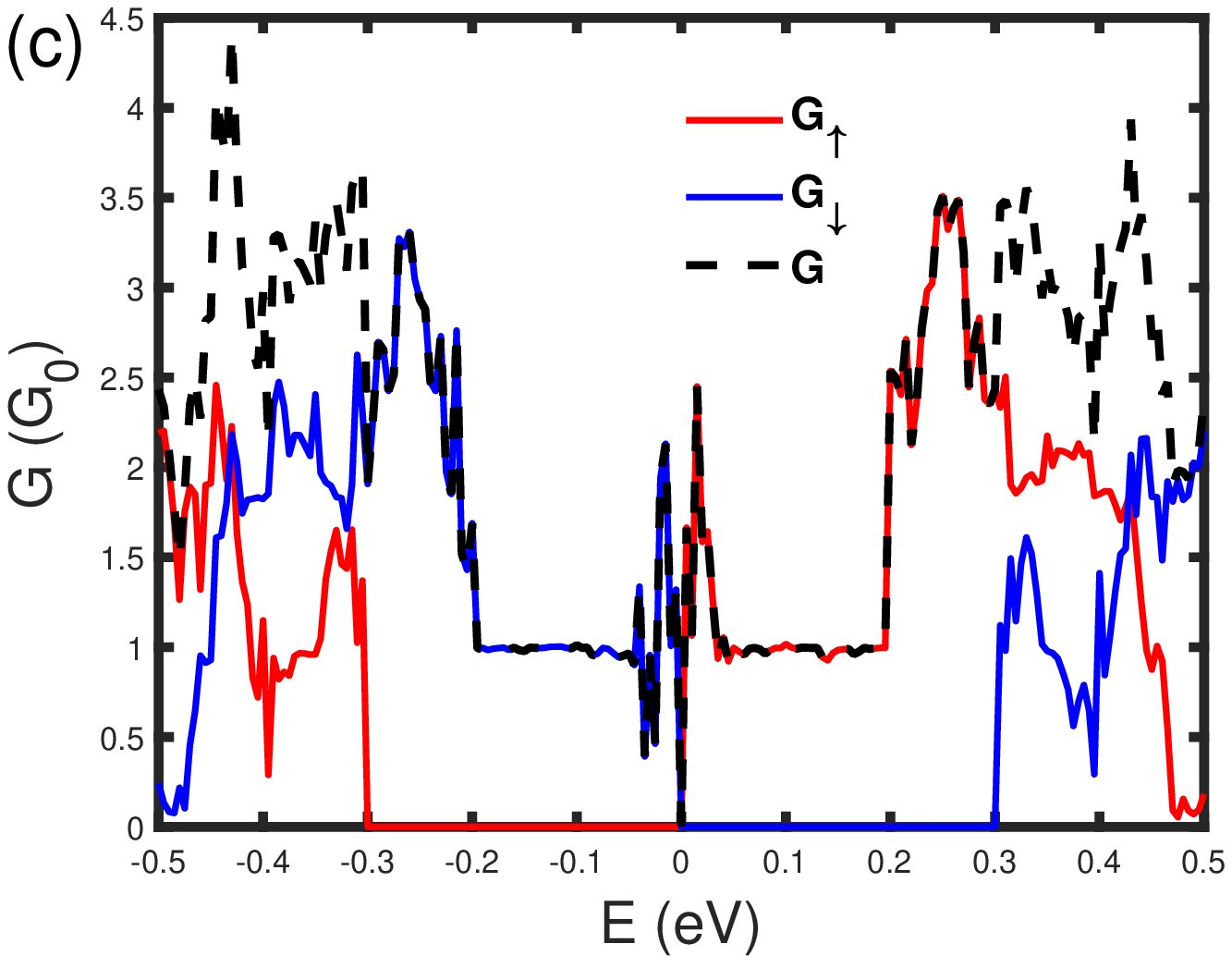}
	\caption{Spin-dependent conductances of two-channel ATSNR for (a) terminal 2, (b) terminal 3 and (c) terminal 4 are plotted  as a function of Fermi energy. Gate voltages of two switches are set to be $V_{1s}=0$,  $V_{2s} = 4$ eV. Other parameters are the same as those in Fig.~\ref{fig:twoG44}.}
	\label{fig:twoG}
\end{figure}

\begin{figure}[t]
	\centering
    \includegraphics[scale=0.4,trim=0 0 0 0,clip]{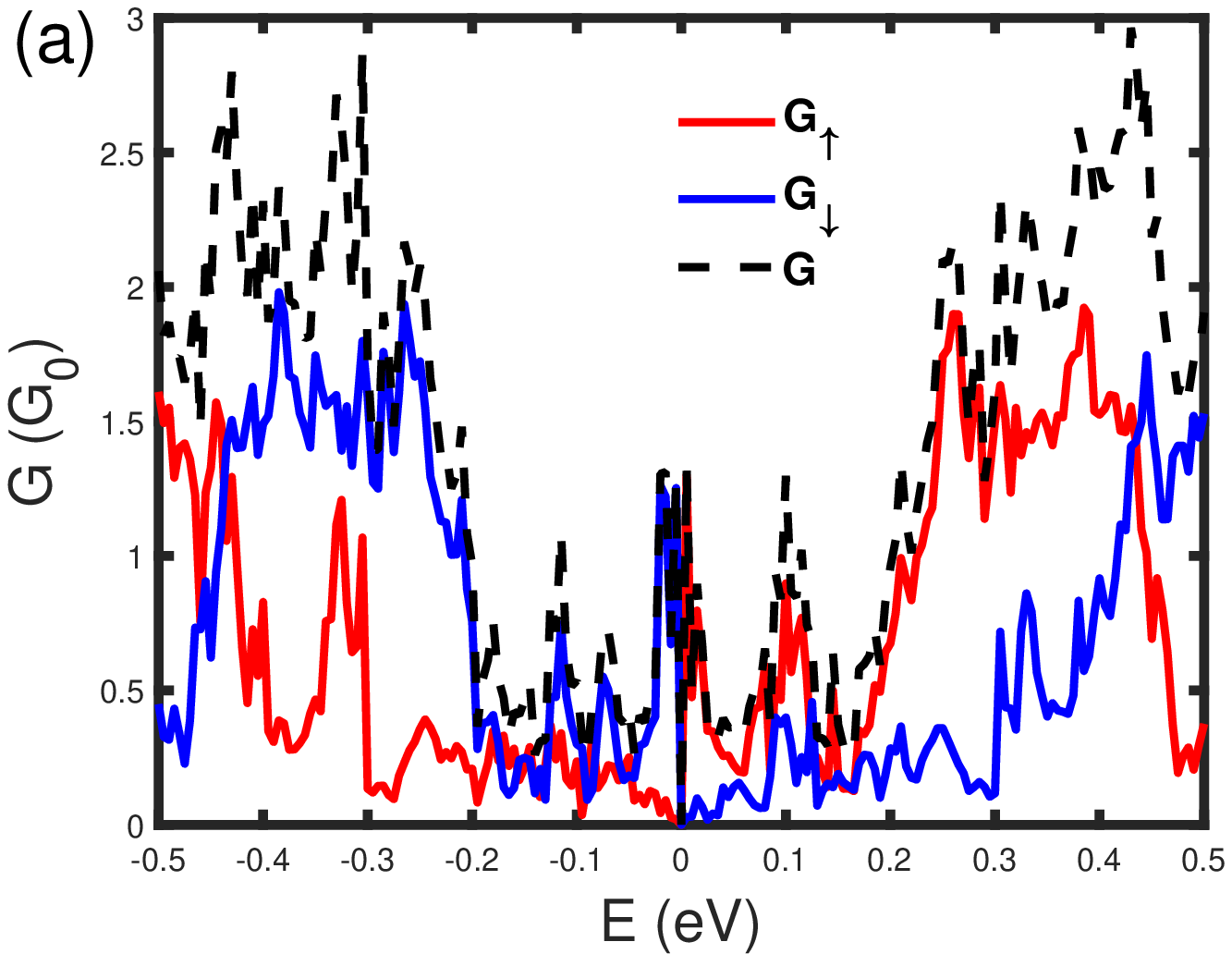}
    \includegraphics[scale=0.4,trim=0 0 0 0,clip]{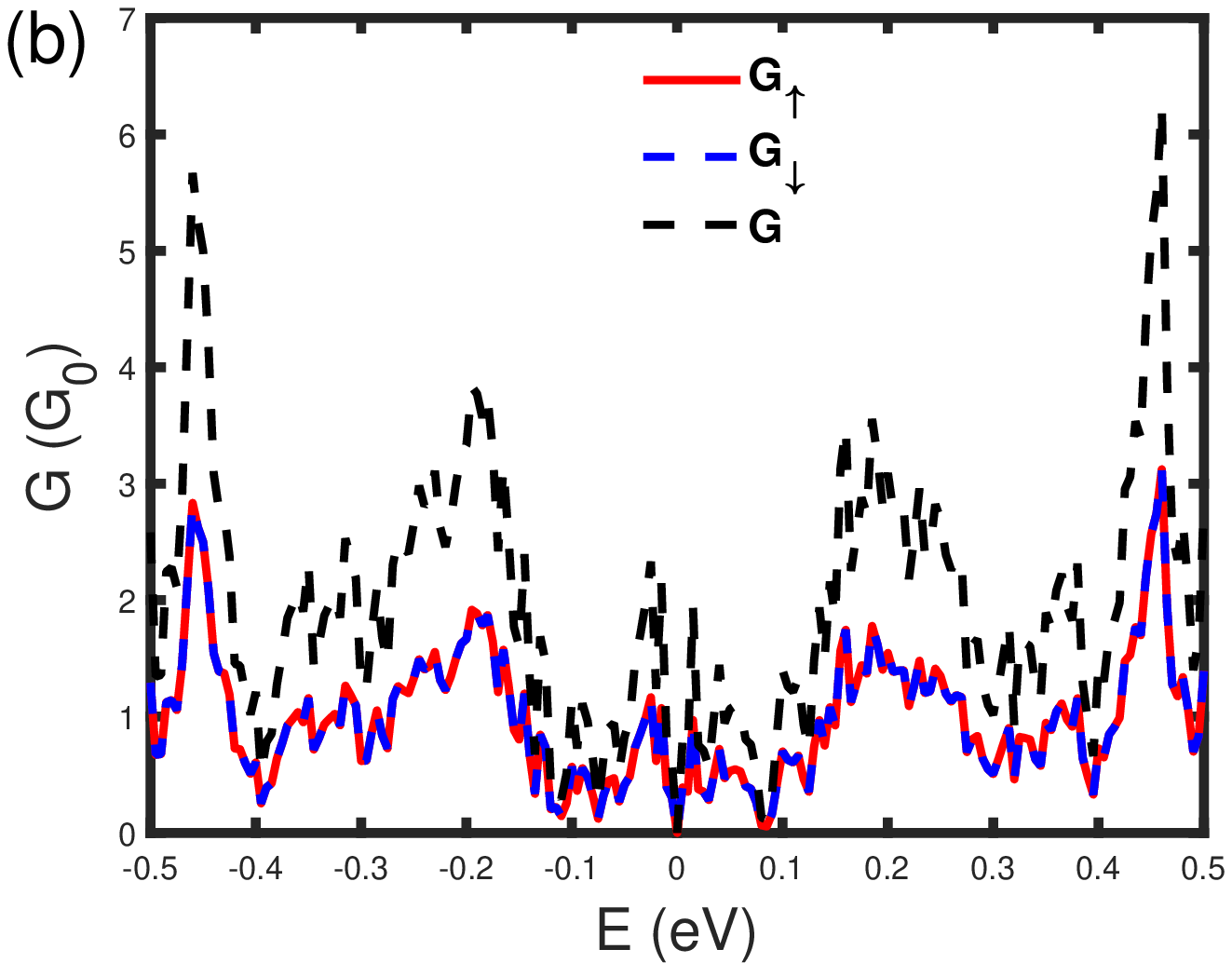}
    \includegraphics[scale=0.4,trim=0 0 0 0,clip]{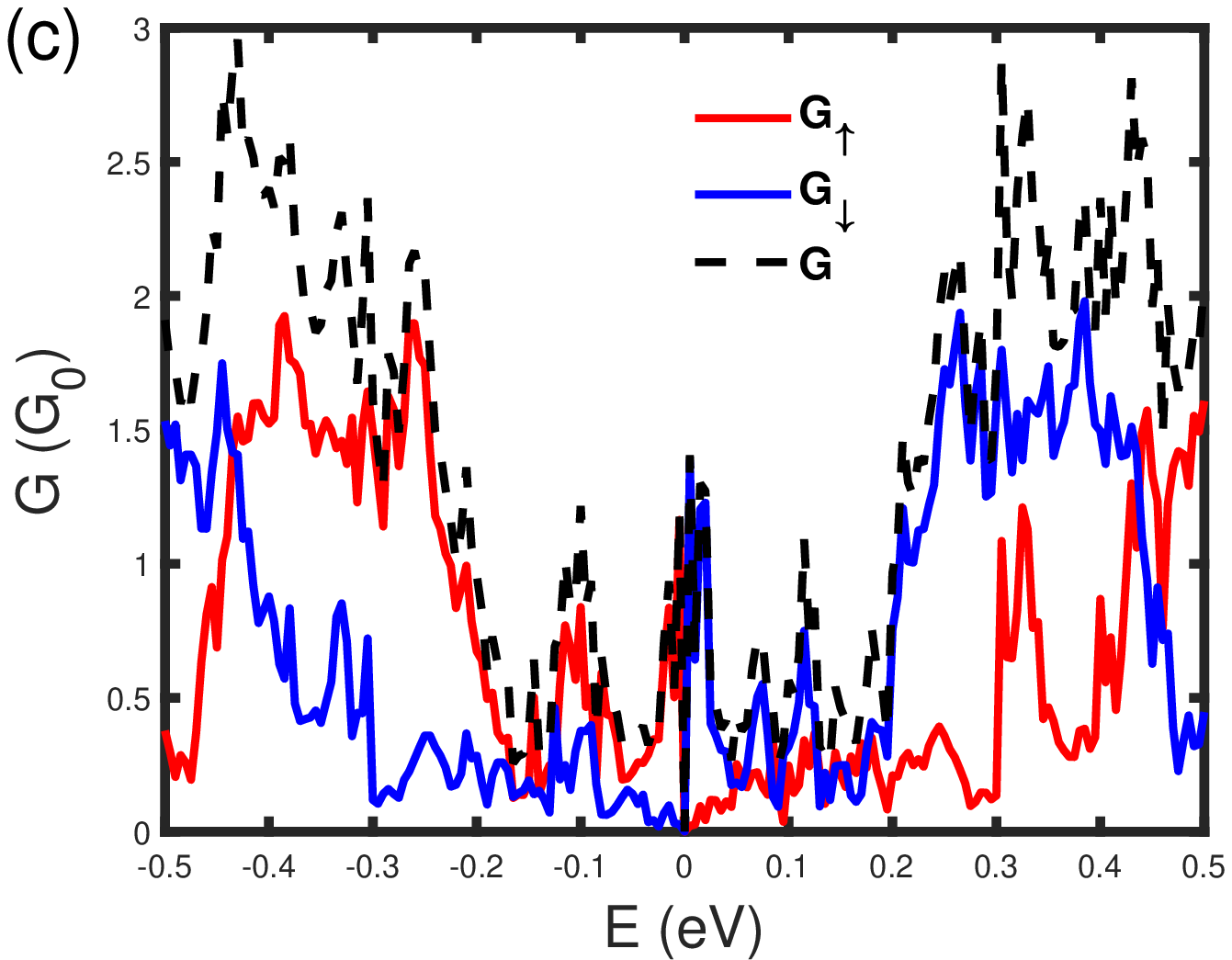}
	\caption{Spin-dependent conductances of two-channel ATSNR for (a) terminal 2, (b) terminal 3 and (c) terminal 4 are plotted  as a function of Fermi energy. Gate voltages of two switches are set to be $V_{1s}=V_{2s}=0$. Other parameters are the same as those in Fig.~\ref{fig:twoG44}.}
	\label{fig:zeroG}
\end{figure}
For terminal 2, we can find that the spin-up conductance $G_\uparrow$ is zero in the energy range $0\leq E\leq 0.3~\mathrm{eV}$, and $G_\uparrow$ gradually increases with oscillating behaviors when $E> 0.3~\mathrm{eV}$. While the spin-down conductance $G_\downarrow$ has a zero value in the energy range $-0.3~\mathrm{eV}\leq E\leq 0$. The spin-dependent zero conductances associate with energy bands given in Fig.~\ref{fig:spinband1}(c). When $M_1=E_{\xi 1}=0.15$eV, there exists a band gap with 0.3eV for spin-up electrons in the energy range $0\leq E\leq 0.3~\mathrm{eV}$, the spin-down band gap occurs at
the energy range $-0.3~\mathrm{eV}\leq E\leq 0$. When $0< E< 0.2~\mathrm{eV}$, spin-dependent subbands evolve from the flat band to fully occupied bands
for all of wave vectors, which result in a spin-down conductance plateaus in this energy range. Similarly, there also exists a conductance plateaus for spin-up electrons in the energy range $-0.2~\mathrm{eV}\leq E\leq 0$. Obviously, the total conductance has no zero value. According to previous analyses, we know that
a band gap can be opened when the amplitude of the sublattice potential is larger than that of the exchange field, namely $E_\xi> M$. It means that one can modulate the total conductance to zero value by increasing the sublattice potential in the two-channel ATSNR. Since channel 2 has opposite amplitudes of the exchange field and the sublattice potential, namely $M_2=E_{\xi 2}=-0.15$eV, the spin-dependent conductances have opposite transport behaviors, as shown in Fig.~\ref{fig:twoG44}(c). Interestingly, we can realize an effective spin polarization by using the two-channel ATSNR. The fully spin currents can be generated and separated into two different terminals in the energy range $-0.3~\mathrm{eV}\leq E\leq 0.3~\mathrm{eV}$.

Next the gate voltage $V_{1s}$ of switch 1 is changed to zero, and $V_{2s}$ keeps 4 eV so as to control the spin current flowing from channel 2.
Seen from Fig.~\ref{fig:twoG}(c), the spin-dependent conductances of terminal 4 basically keep unchanged due to high potential barrier of switch 2.
However, when $V_{1s}=0$, the maximum value of conductances of terminal 2 significantly decreases to about $1.32 G_0$ and the conductance plateaus also disappears [see Fig.~\ref{fig:twoG}(a)]. It is obvious that the conductances of terminal 3 significantly increase and zero-conductance ranges are identical with those of terminal 2. But the conductance amplitude of terminal 3 is larger than that of terminal 2. It means that electrons prefer entering into terminal 3 through switch 1. When $V_{1s}=4$eV and $V_{2s}=0$, namely switch 2 is open and switch 1 is closed, similarly, electrons will flow from channel 2 into terminal 3 through switch 2. Correspondingly, the transport behaviors of terminals 2 and 4 are exchanged comparing with previous switch configuration.

\begin{figure}[t]
	\centering
    \includegraphics[scale=0.4,trim=0 0 0 0,clip]{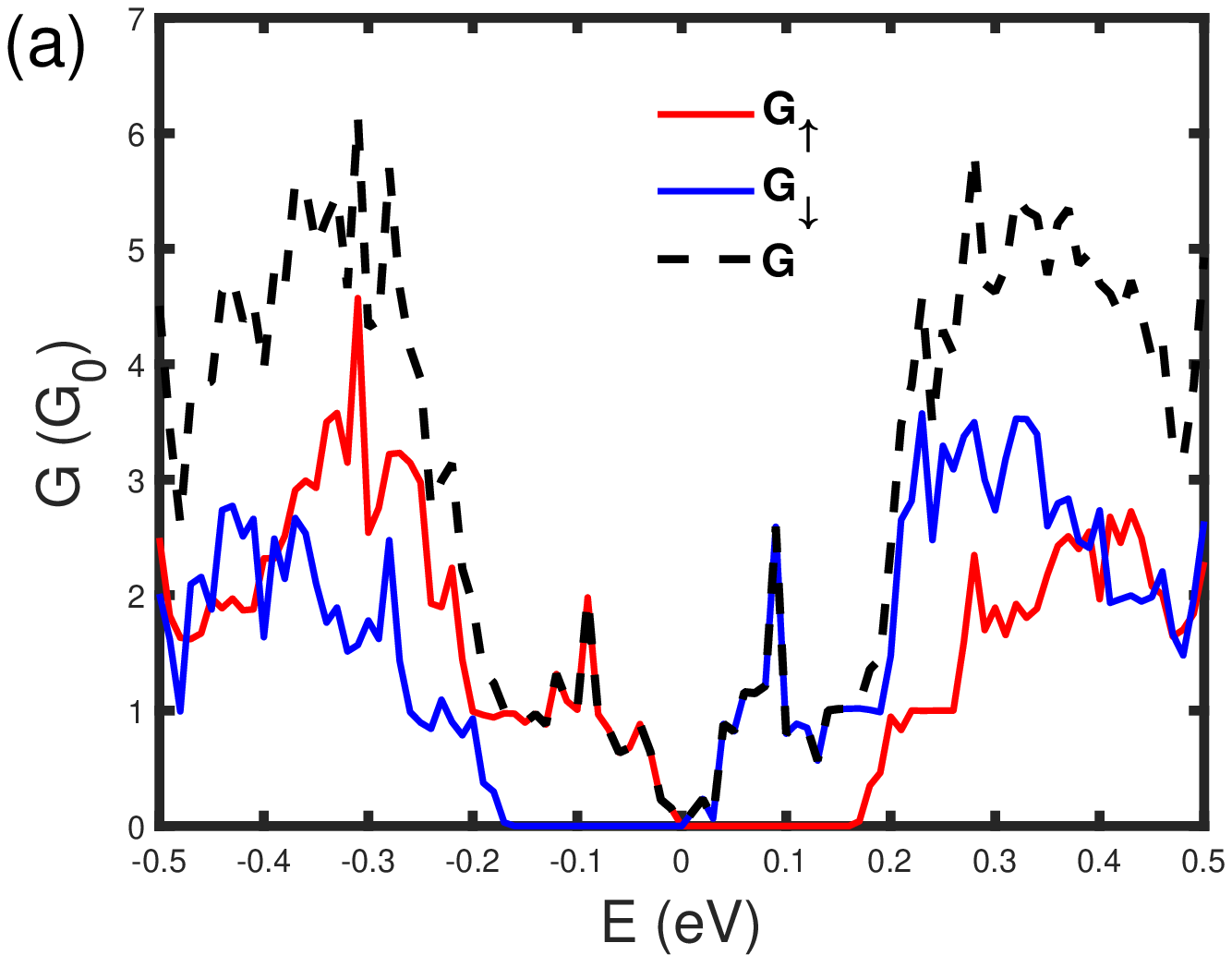}
    \includegraphics[scale=0.4,trim=0 0 0 0,clip]{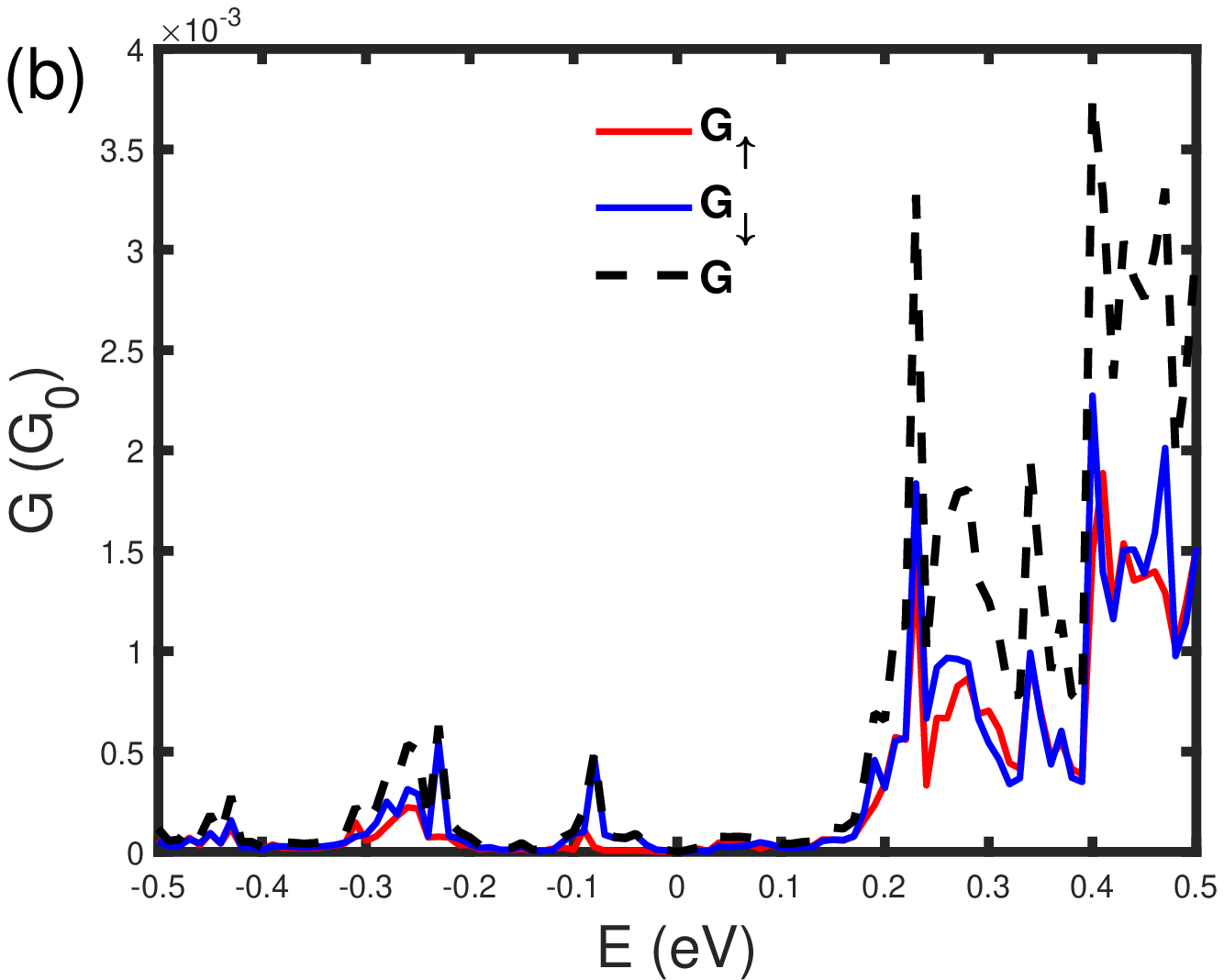}
    \includegraphics[scale=0.4,trim=0 0 0 0,clip]{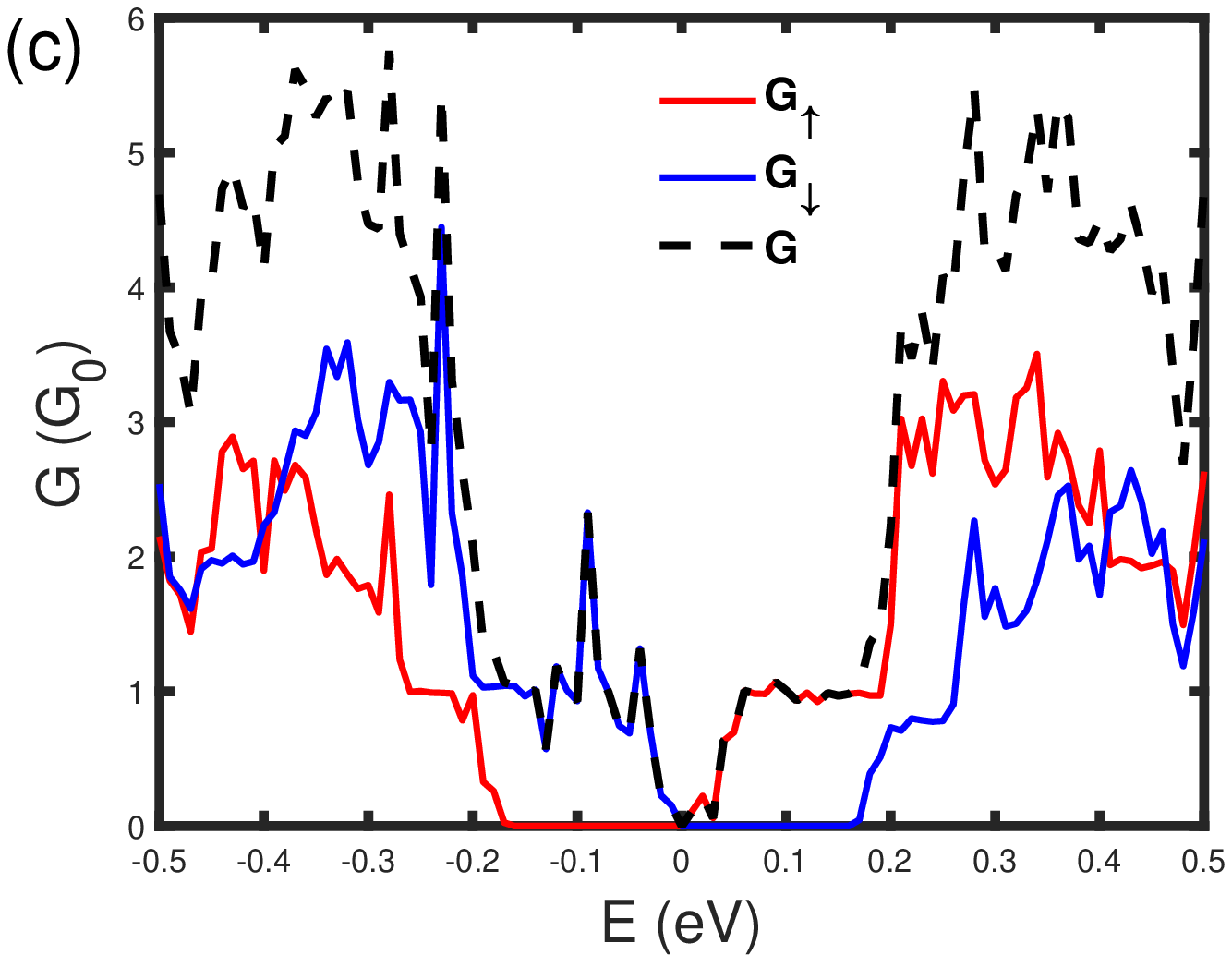}
\caption{Spin-dependent conductances of two-channel ATSNR for (a) terminal 2, (b) terminal 3 and (c) terminal 4 are plotted  as a function of Fermi energy. Gate voltages of two switches are set to be $V_{1s} = V_{2s} = 4$ eV. The spin-orbit coupling is $t_{so}=0.1$ eV, other parameters are given by $E_{\xi 1}=0.15$ eV and $ M_1 = 0.084$ eV in channel 1, $E_{\xi 2}=-0.15$ eV and $ M_2 = -0.084$ eV in channel 2.}
	\label{fig:twoG44SOC}
\end{figure}

\begin{figure}[t]
	\centering
    \includegraphics[scale=0.4,trim=0 0 0 0,clip]{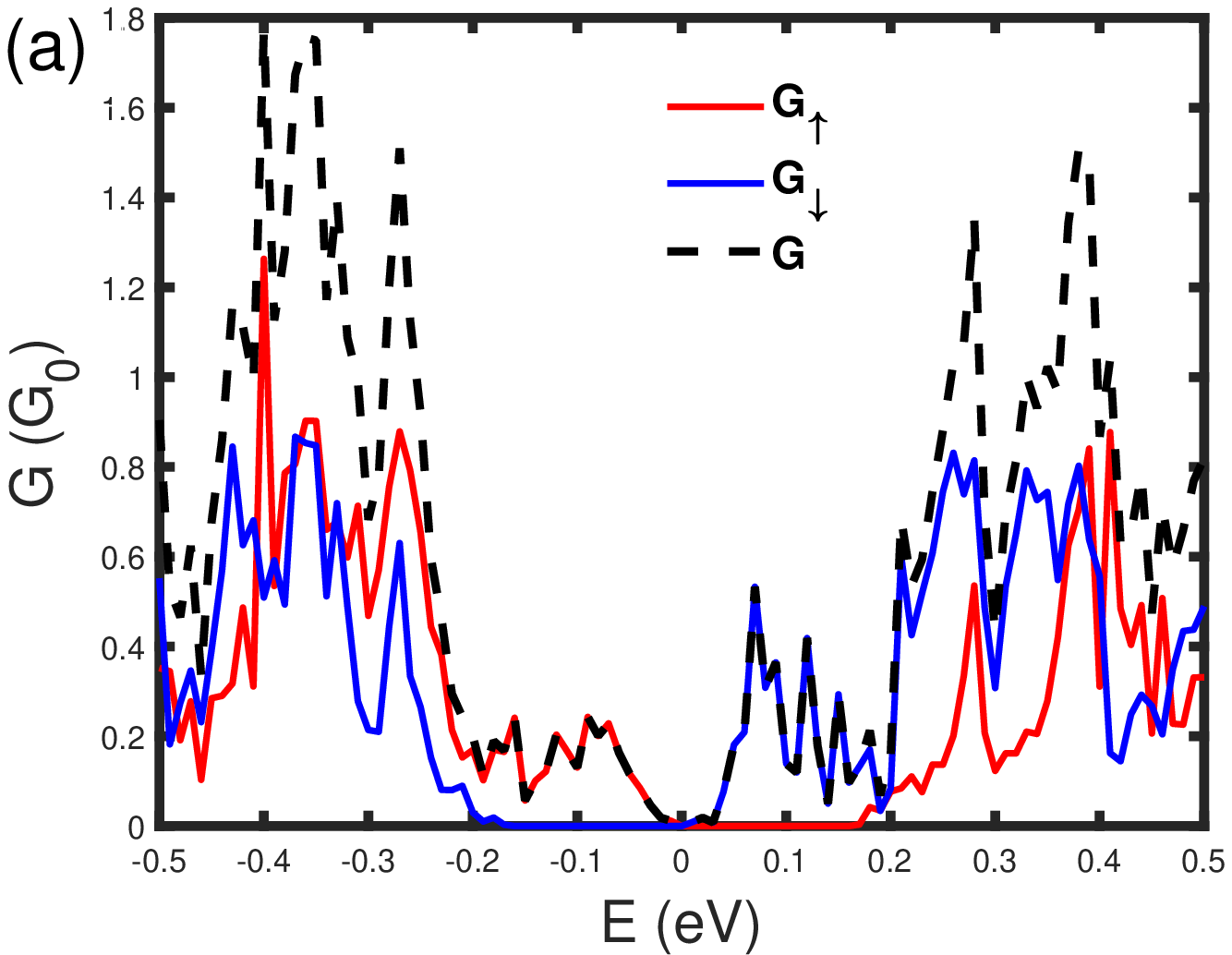}
    \includegraphics[scale=0.4,trim=0 0 0 0,clip]{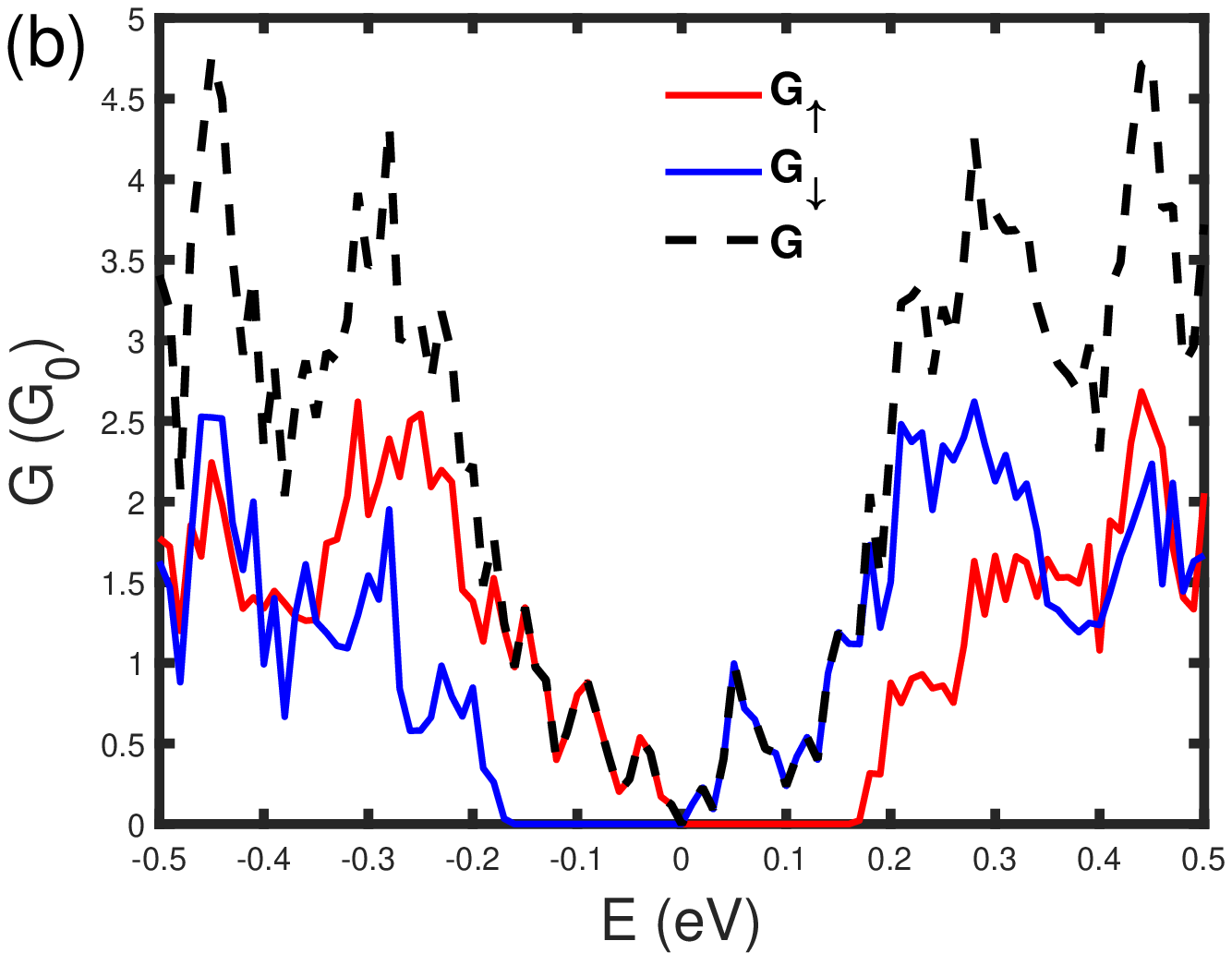}
    \includegraphics[scale=0.4,trim=0 0 0 0,clip]{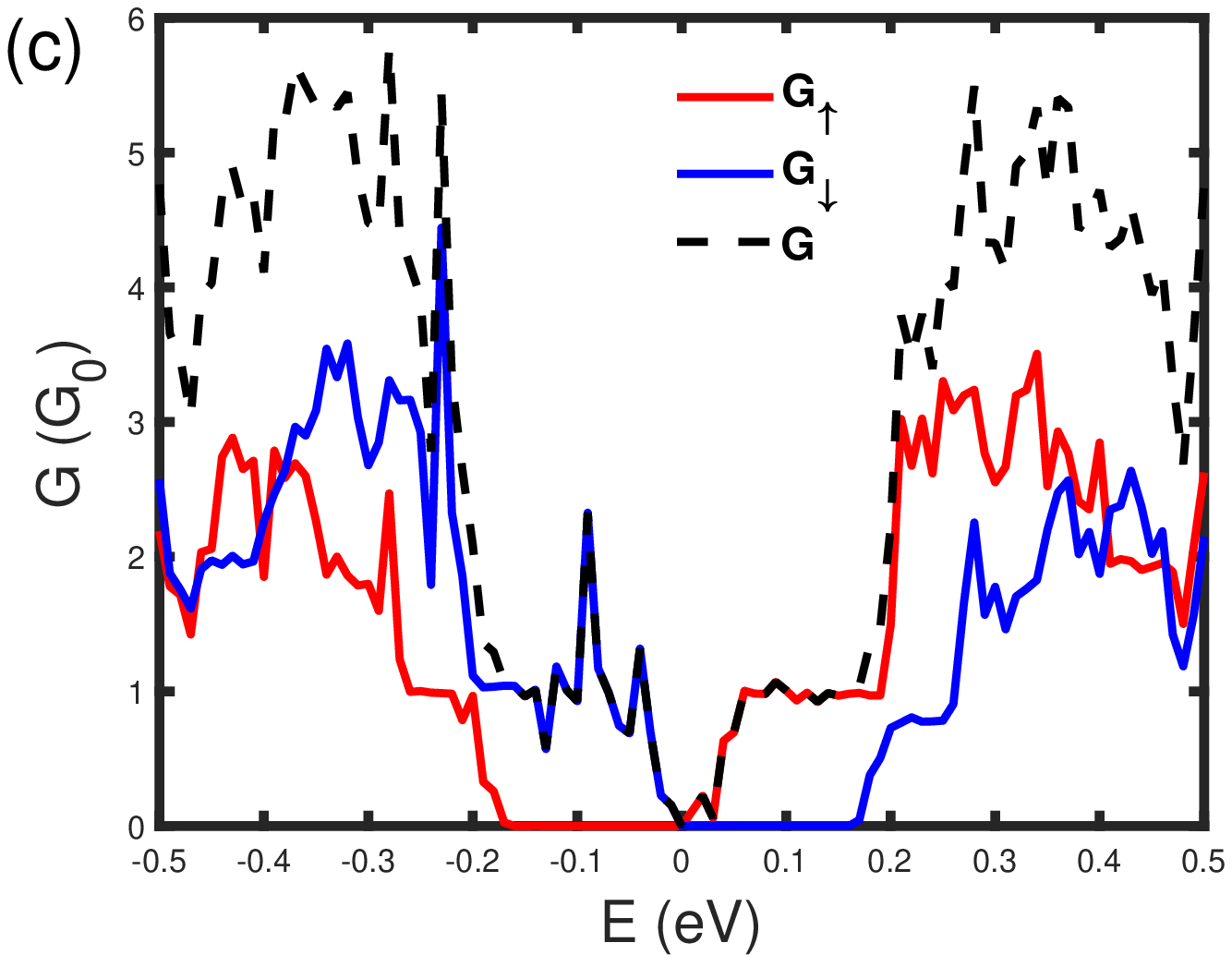}

	\caption{Spin-dependent conductances of two-channel ATSNR for (a) terminal 2, (b) terminal 3 and (c) terminal 4 are plotted  as a function of Fermi energy. Gate voltages of two switches are set to be $V_{1s}=0$, $V_{2s} = 4$ eV. The spin-orbit coupling is $t_{so}=0.1$ eV, other parameters are the same as those in Fig.~\ref{fig:twoG44}.}
	\label{fig:twoG04SOC}
\end{figure}
When the two switches are open, namely $V_{1s}=V_{2s}=0$, the zero-conductance ranges for terminals 2 and 4 disappear, as shown in Fig.~\ref{fig:zeroG}.
We can see that spin-up conductance $G_\uparrow$ of terminal 2 is larger than $G_\downarrow$, which is zero in the energy range
$0\leq E\leq 0.3~\mathrm{eV}$ for other switch configurations. The conductances of terminal 4 have reversed spin-dependent phenomenon.
Interestingly, the spin-dependent conductances of $G_\uparrow$ and $G_\downarrow$ are identical in terminal 3. It means that there exists no spin polarization in terminal 3. In terms of above analyses, spin-dependent conductance can be effectively modulated by utilizing one switch or two switches. Especially,
applying only one switch can realize a spin-dependent zero conductance and large spin polarization. The amplitudes of the exchange field and the sublattice potential are crucial for creation of spin currents. Then two switches can provide multiple operations for controlling of spin-dependent transport properties.

\subsection{Generation of fully spin-polarized current via two-channel device in the presence of SOC}
Now we start to study the combined effect of the spin-orbit coupling, exchange field and the sublattice potential on the spin-dependent conductance. The amplitude of SOC is set to be $t_{so}=0.1$ eV. Gate voltages of the two switches are set to be $V_{1s} = V_{2s} = 4$ eV. The two channels have opposite amplitudes of exchange fields and sublattice potentials, namely $E_{\xi 1}=0.15$ eV and $ M_1 = 0.084$ eV in channel 1, $E_{\xi 2}=-0.15$ eV and $ M_2 = -0.084$ eV in channel 2. The spin-dependent conductances of three terminals of the two-channel ATSNR are plotted  as a function of Fermi energy, as shown in Fig.~\ref{fig:twoG44SOC}.
Seen from Fig.~\ref{fig:spinband1}, there should exist a band gap of 0.132 eV for the case $E_{\xi 1}=0.15$ eV and $ M_1 = 0.084$ in the absence of SOC. While SOC induces tilted subbands and results in disappearance of the band gap [see Fig.~\ref{fig:SOCband2}]. Thus the total conductance has no zero value in Fig.~\ref{fig:twoG44SOC}.
It is found that the spin-up conductance of terminal 2 has a zero value in the energy range $0<E<0.18 \mathrm{eV}$, while the spin-down conductance has a zero value in the energy range $-0.18 \mathrm{eV}<E<0$, which associate with the tilted subbands. Obviously, the spin-dependent conductances are not symmetrical about the axis $E=0$ because of the tilted subbands, which are different from the cases without SOC. Similarly, the spin-dependent conductances of terminal 4 have opposite transport behaviors due to the opposite amplitudes of $E_{\xi 2}$ and $M_2$. Certainly, the conductances of terminal 3 are controlled to be small values by the action of the high potential barriers of $V_{1s}$ and $V_{2s}$.

When switch 1 is opened, namely $V_{1s}=0$, the conductances of terminal 3 significantly increase, while the conductances of terminal 2 accordingly decrease. The maximum value of the total conductance in terminal 3 is about $4.7G_0$. What's more, the spin polarization of terminal 3 is similar with that of terminal 2. It means that the spin-polarized electrons flow from channel 1 into terminal 3 via switch 1. The conductances of terminal 4 keep the same variation trends as those given in Fig.~\ref{fig:twoG44SOC}(c) due to the high potential of switch 2. It is obvious that one can still obtain the spin-polarized conductance in the presence of SOC, even though the zero-conductance ranges are different.
When the two switches are open, namely $V_{1s}=V_{2s}=0$, the zero-conductance ranges of terminals 2 and 4 disappear.
The spin-dependent conductances of $G_\uparrow$ and $G_\downarrow$ are basically identical in all of the three terminals. It means that the spin polarization closes to zero. In terms of above analyses, SOC can induce tilted energy bands and modulate the size of the band gap. Accordingly, the zero-conductance range and the spin polarization can be affected by SOC. As we expected, the two-channel ATSNR can realize an effective separation of spin current. This phenomenon suggests that tetragonal silicene may be a potential candidate for spintronic devices.

\section{CONCLUSIONS}\label{sec:Conclusions}
In summary, we have studied the band structure and transport properties of ferromagnetic tetragonal silicene nanoribbon by using non-equilibrium Green's function method. The band structure and spin-dependent conductance are discussed under the combined effect of the external electric field (sublattice potential), potential energy, the exchange field and SOC. The dispersion relation of ATSNR depends on the transverse width of the nanoribbon. One can easily realize a phase transition from a semimetalic to semiconducting state by changing the transverse width of the nanoribbon. A band gap of tetragonal silicene can be opened due to the effect of the electric field. The combined effect of the exchange field and electric field can result in spin-dependent band gaps.
SOC can induce tilted subbands and modulate the band gap. The tilted angle gradually increases with increasing of the strength of SOC.
Separation of spin-dependent conductances arises from the effect of exchange field and SOC, while zero-conductance behaviors exhibit spin-dependent band gaps induced by the electric field. In terms of the band structure and spin-dependent conductance, we propose a device configuration of four-terminal two channels spin-based ATSNR. The central two channels work under the combined effect of the exchange field, sublattice potential and SOC. The drain lead is divided into three terminals connecting with right buffer layer. We find that the spin current can be controlled by utilizing two switches. The switch with high potential barrier can block electrons flowing from central scattering region into other terminals. Interestingly, applying only one switch can realize spin-dependent zero conductance and large spin polarization. The amplitudes of the exchange field, sublattice potential and SOC are crucial for generation of spin currents. Two switches can provide multiple operations for controlling spin-dependent transport properties. The two-channel ATSNR can realize an effective separation of spin current, which may be a potential candidate for spintronic devices.

\begin{acknowledgments}
This work was supported by National Natural Science Foundation of China (Grant No. 11574067).
\end{acknowledgments}

\end{document}